\documentclass{aa}
\usepackage{graphicx, natbib,epsfig,amsmath,amssymb, mathrsfs, txfonts}
\DeclareGraphicsExtensions{.eps, .jpg, .ps}
\bibpunct{(}{)}{;}{a}{}{,}
\begin{document}

\title{Comparison of coronagraphs for high contrast imaging in the context of Extremely Large Telescopes}
\author{P. Martinez\inst{1, 2, 6}  \and A. Boccaletti\inst{2, 6} \and M. Kasper\inst{1} \and C. Cavarroc\inst{3} \and N. Yaitskova\inst{1} \and  T. Fusco\inst{4, 6} \and  C. V\'{e}rinaud\inst{5}}
\institute{European Southern Observatory, Karl-Schwarzschild-Strasse 2, D-85748, Garching, Germany 
\and LESIA, Observatoire de Paris Meudon, 5 pl. J. Janssen, 92195 Meudon, France
\and CEA, Saclay, 91191 Gif-sur-Yvette cedex, France
\and ONERA, BP 52, 29 avenue de la division Leclerc, 92320 Chatillon cedex, France
\and LAOG, Observatoire de Grenoble, 38041 Grenoble, France
\and Groupement d\'{}int\'{e}r\^{e}t scientifique PHASE (Partenariat Haute r\'{e}solution Angulaire Sol Espace)}
\offprints{P. Martinez, martinez@eso.org}
\abstract
{}
{We compare coronagraph concepts and investigate their behavior and suitability for planet finder projects with Extremely Large Telescopes (ELTs, 30-42 meters class telescopes).} 
{For this task, we analyze the impact of major error sources that occur in a coronagraphic telescope (central obscuration, secondary support, low-order segment aberrations, segment reflectivity variations, pointing errors) for phase, amplitude and interferometric type coronagraphs. This analysis is performed at two different levels of the detection process: under residual phase left uncorrected by an eXtreme Adaptive Optics system (XAO) for a large range of Strehl ratio and after a general and simple model of speckle calibration, assuming common phase aberrations between the XAO and the coronagraph (static phase aberrations of the instrument) and non-common phase aberrations downstream of the coronagraph (differential aberrations provided by the calibration unit).}
{We derive critical parameters that each concept will have to cope with by order of importance. 
We evidence three coronagraph categories  as function of the accessible angular separation and proposed optimal one in each case.
Most of the time amplitude concepts appear more favorable and specifically, the Apodized Pupil Lyot Coronagraph gathers the adequate characteristics to be a baseline design for ELTs.}
{}


\maketitle


\section{Introduction}

Recent years have seen intensive research and development of new high contrast imaging technics that are essential for detecting faint structures or companions around bright parent stars. 

A variety of astrophysical topics (low-mass companions, circumstellar disks, ...) has driven the next generation of high contrast instruments like SPHERE and GPI \citep{2006Msngr.125...29B,2006SPIE.6272E..18M} expected in 2011, or EPICS \citep{EPICS} for the longer term ($\sim$ 2018). Coronagraphy is a mandatory technique for these instruments and is therefore a critical sub-system. 

A large review of the different families of coronagraph was carried out by \citet{2006ApJS..167...81G} and optimal concepts were proposed in the context of space-based observations. Results of this study cannot be generalized for ground-based observations as the problematic is different. Contrast level requirements are relaxed while telescopes 
parameters may have a different impact.

We have previsously studied the contrast performance of Extremely Large Telescopes (ELTs) and limitations for an "ideal" coronagraph \citep{2006A&A...447..397C} and in \citet{2007A&A...474..671M}, we  have shown how the Apodized Pupil Lyot Coronagraph (APLC) can be optimized with respect to the ELT parameters. 
Here,  the objective is to investigate the trade-off for coronagraphy in the general context of  ELTs. Telescope characteristics such as central obscuration ratio, primary mirror segmentation, and secondary mirror supports can have an impact on high contrast imaging capabilities and impose strong limitations for many coronagraphs. Coronagraphs already selected for 8-10 m class telescopes are not necessary suited for future planet finder projects with 30-42 m ELTs for which the achievable angular resolution becomes extremely high ($\sim$ 10 mas).

On ground-based telescopes equipped with extreme Adaptive Optics systems (XAO), coronagraphs are expected to attenuate significantly the on-axis star. However, even at a high level of correction (Strehl ratio $>$ 90\%) a significant fraction of the star flux remains in the focal plane ($<$10\%). The residual light sets the photon noise contribution for high contrast imaging, even if a a dedicated calibration procedure like differential imaging is in used \citep{1999PASP..111..587R, 2000PASP..112...91M, 2003PASP..115.1363B, 2004ApJ...615..562G}. The estimation of this level is thus one byproduct of our study (as shown in  \citealt{2006A&A...447..397C})

The intent of this paper is twofold: 1/ determine limiting parameters and ideally derive specifications at the system level, 2/ initiate a general comparison of coronagraphs to identify valuable concepts and field of application. 

In Sect. \ref{sec:methodology} all the simulation hypothesis are described: coronagraph concepts, AO and calibration unit assumptions, metrics used and error sources considered. The impact of these error sources with respect to the AO correction level and their effects on the detectability using a differential imaging system are presented in Sect. \ref{sec:study1} and discussed in Sect. \ref{sec:study2}. Further comparisons between two promising concepts is done in Sec. \ref{compAPLCBL4}. Finally, in Sect. \ref{sec:conclusion} we derive conclusions. 

\section{Methodology}
\label{sec:methodology}

\subsection{Coronagraphs}
For the last ten years, a wide number of coronagraph concepts has been proposed. Nevertheless, at this time, none of them can simultaneously meet all of the main high contrast requirements:  not being sensitive to telescope parameters (like the secondary support obscuration, residual aberrations, spider vanes), small inner working angle (IWA), not being sensitive to pointing errors, high throughput, good imaging capabilities (large searchable area), compatible with large bandwidth, and finally, manufacturing feasibility.
The effect of the segmentation of the primary mirror (inter-segment gaps, segment reflectivity variations, segment aberrations) must be added to this long list as we are dealing with ELTs. 

\subsubsection{Concepts analyzed in this study}
We consider, the following coronagraph concepts: Lyot coronagraph \citep{1939MNRAS..99..580L}, Apodized Pupil Lyot Coronagraph \citep[APLC,][]{2002A&A...389..334A, 2003A&A...397.1161S}, Apodized Roddier \& Roddier Coronagraph (i.e Dual zone) \citep[APRC,][]{2003A&A...397.1161S, 2003A&A...403..369S}, Four Quadrant Phase Mask \citep[FQPM,][]{2000PASP..112.1479R, 2001PASP..113.1145R}, Annular Groove Phase Mask \citep[AGPM,][]{2005ApJ...633.1191M}, Band-limited \citep[BL,][]{2002ApJ...570..900K, 2005ApJ...628..466K}, Achromatic Interferometric Coronagraph \citep[AIC,][]{1997ASSL..215..187G,2000A&AS..141..319B, 2000A&AS..145..341B, 2005PASP..117.1004B}. 

In most figures presented in this paper, we will be make use of an ideal coronagraph model that removes the flat non-tilted coherent wavefront from the pupil \citep{2006A&A...447..397C}. In such a way, the ideal coronagraph enables to derive the maximum detection level imposed either by the residual phase aberrations left from the XAO system or the static aberrations from the differential imaging system (as defined in next sections). Ideal coronagraph is only designed for no wavefront aberrations and comparison to other coronagraphs is useless.

\subsubsection{Inner working angle constraint}
\label{sec:IWA}
The IWA describes quantitatively how close a coronagraph design allows the detection of a faint companion. In this paper we define the IWA as the angular separation for which the diffraction peak of a planet is reduced by a factor of 2. 
It assumes that the planet's throughput is radially averaged (i.e not favorably placed) while the star is point-like.\\
The AIC, FQPM/AGPM, APRC have a very small IWA owing to their intrinsic properties.
On the opposite, amplitude concepts (Lyot, APLC, and BLs) have a larger
IWA depending on coronagraph parameters (d, diameter of the focal mask or $\epsilon$, bandwidth of the mask function that actually depends on the application). 

Since we are dealing with ELTs, the angular resolution of such large telescopes is relaxing the constraint on the IWA and hence the problematic is different than for planet finder instruments on 8-m class telescopes. 
As a baseline, we fixed the limit of the IWA to the reasonable value of 4$\lambda/D$ . For instance, at 1.6 $\mu m$ (H-band), 4$\lambda/D$ correspond to 30 mas and 165 mas for a 42 and a 8 meters telescope respectively. 

In the next simulations, the Lyot coronagraph has a mask size of 7.5$\lambda/D$ (i.e a corresponding IWA of 3.9 $\lambda/D$). 
The APLC has a 4.7$\lambda/D$ mask diameter (i.e IWA = 2.4 $\lambda/D$). This size is the result of the optimization performed in \citet{2007A&A...474..671M}.
We also consider two band-limited masks with different orders: a $4^{th}$ order \citep[BL4,][$sin^{4}$ intensity mask with $\epsilon$ = 0.21]{2002ApJ...570..900K} and a $8^{th}$ order \citep[BL8,][m=1, l=3 and $\epsilon$ = 0.6]{2005ApJ...628..466K}. BLs parameter $\epsilon$ both control the IWA and Lyot stop throughput. 

\subsubsection{Pupil stop optimization}
The total amount of the rejected light by a coronagraph strongly depends on the pupil stop size and shape. In this paper, pupil stops are optimized to match the diffraction in the relayed pupil as defined in \citet{2004EAS....12..165B} and hence are well adapted to the way that each coronagraph deal with the diffracted light.
However, in Sec. \ref{sec:study1} we generate a large different range of wavefront errors, hence an optimization of the pupil stop with respect to the level of the residual phase could relax constraints on the pupil stop shape and throughput (as discussed for instance in \citealt{2007ApJ...661.1323C}, for the Band-Limited case). This optimization depends on the dominant source of noise (diffracted light or uncorrected atmospheric speckles).

In practice, we optimized pupil stops in the ideal case (no wavefront error) since the final comparison is made after differential imaging when the uncorrected atmospheric speckles have been removed. Pupil stops are assumed to be perfectly aligned except when we evaluate the impact of its misalignment (pupil stop throughput and coronagraph parameters are summarized in Table \ref{coro1}).

\subsection{Principle of numerical simulations}
As a baseline we consider a 42 meters ELT with 30\% (linear) central obscuration ratio as expected for the European ELT (E-ELT, \citet{E-ELT2008}), except when we evaluate its impact. As for the wavelength, we adopt a baseline of $\lambda=1.6\mu m$ (centre of the H-band), a good compromise between angular resolution and AO correction. Our simulations make use of simple Fraunhofer propagators between pupil and image planes, which is implemented as fast Fourier transforms (FFTs) generated with an IDL code. The image plane is sampled with 0.125$\lambda/D$ per pixel.

\subsubsection{Extreme AO simulations}
Since we are concerned with ELTs, we consider an eXtreme Adaptive Optics system (XAO) with a large number of actuators. Table \ref{parameter} shows the characteristics for the simulations of the XAO phase residuals. As we want to analyze the behavior of coronagraph under realistic conditions we generate many phase screens with different Strehl ratios (from 84\% to 96\%). For that, we modify the atmospheric seeing (from 1.0\arcsec to 0.4\arcsec) while leaving the XAO system unchanged. As a fair comparison, all coronagraphs have been affected by the same set of phase screens. We are using 100 phase realizations, and check that it was sufficient to produce long exposures at the contrast level we are dealing with. 

\subsubsection{Differential imaging simulations}
\label{DIsimul}
The presence of a residual atmospheric wavefront perturbation even if corrected with a XAO system is limiting the contrast behind a coronagraph to about $10^4$ - $10^6$. 
\begin{table*}[!ht]
\centering
\caption{Parameters of coronagraphs optimized for a central obscuration of 30\%. d is the Lyot focal mask diameter, $\epsilon$ the BL bandwidth parameter (m and l are complementary BL8 function parameters), lp is the AGPM topological charge and $\mathscr{T}$ the overall transmission. The last column gives the status of coronagraph developments for lab experiment as well as on-sky observations.}
\label{coro1}
\begin{tabular}{l|l|l|l|l|l}
\hline \hline 
Coronagraph type &\multicolumn{3}{c|}{Specifications} & \multicolumn{2}{c}{State of art}   \\
\cline{2-6} \\
& IWA ($\lambda/D$, $\pm$ 0.1) &$\mathscr{T}$ (\%) & Parameters &  Laboratory tests  & Sky observations \\
\hline
FQPM & 0.9 & 82.4   & -  & vis. / near IR / mid IR &  near IR \\
AGPM & 0.9 & 82.7 & lp=2 & - & - \\
AIC & 0.4  & 50.0  & -   & vis.  / near IR & near IR \\
Lyot &  3.9 &  62.7   & $d = 7.5 \lambda/D$  &vis. / IR  & vis. / IR  \\
APLC &  2.4 & 54.5   &$d = 4.7 \lambda/D$   & vis. / near IR  & near IR  \\
APRC &  0.7 & 74.5  & $d = 1.06 \lambda/D$ & - &- \\
BL4 & 4.0 & 22.4 & $\epsilon = 0.21$  & vis. & - \\
BL8 &  4.0 &  13.8   & $\epsilon = 0.6$, m=1, l=3  &vis. & - \\
\hline
\end{tabular}
\end{table*}
The first series of simulations to assess the impact of telescope parameters on coronagraph performance is carried out at this level (sect. \ref{sec:XAO}). 
However, it is important to perform the same analysis at the level of contrast which is adequate for planet detection ($10^8$ - $10^{10}$) to evaluate how the sensitivity of coronagraph propagates.
To enhance the contrast, a second step is required to suppress the speckle noise (composed of dynamical and static aberrations). On SPHERE and GPI, specke calibration is implemented in the form of spectral and polarimetric differential imaging \citep{1999PASP..111..587R, 2000PASP..112...91M, 2003PASP..115.1363B}. A larger contrast is then achievable through appropriate data reduction.
Here, for sake of generality we assume a general and simple scheme of Differential Imaging (DI). 
A detailed analysis of contrast performance for ELTs with DI has been performed by \citet{2006A&A...447..397C}. For the reader's convenience we repeat in the following the main assumptions and results used in this present study.
We consider two images taken simultaneously using two channels downstream of the coronagraph (same spectral band, same polarization state). In such a case, the contribution to the wavefront error is made of two terms :
the static common aberrations ($\delta_c$) in the instrument upstream of the coronagraph and non-common aberrations ($\delta_{nc}$) downstream of the coronagraph. The latter corresponds to differential aberrations since the light goes through two different optical paths. 
Here, the residual phase left uncorrected by the XAO system is omitted since it will be averaged to an azimuthally constant pattern over time and be suppressed by subtraction in the two channels (if the photon noise is neglected). Therefore, the detectability for an infinitely long exposure only depends on $\delta_c$ and $\delta_{nc}$. The static aberrations $\delta_c$ and $\delta_{nc}$ are described by PSDs with $f^{-2}$ variation ($f$ is the spatial frequency). Since aberrations are critical at close angular separations,  we assume that the PSDs at low frequencies were "shaped" (flat in the range $0 < f < f_c/4$, with $f_c$ the cut-off frequency of the XAO. 

Many combinations of $\delta_c$, $\delta_{nc}$ are possible to reach the desired contrast level. But, as we are interested in the DI performance rather than the technique itself we adopt in sect. \ref{sec:DI} an arbitrary amplitude of 10 nm rms and 0.3 nm rms for $\delta_c$ and $\delta_{nc}$ respectively. A contrast level of $10^{9}$ is thus achievable which is consistent with EPICS science contrast requirements \citep{EPICS}.
\begin{center}
\begin{table*}
\centering
\caption{Values and amplitudes of parameters used in the simulation.}
\label{parameter}
\begin{tabular}{l|l|l}
\hline \hline
 & XAO simulation & DI simulation \\
\hline 
\textbf{Input parameters} & & \\
\hline 
telescope diameter & 42 m & 42 m \\
seeing at 0.5 $\mu$m & $1.0\arcsec - 0.4 \arcsec$ & -\\
wind speed & 15 m/s & -\\
outerscale of turbulence $L_{0}$ & 20 m & -\\
number of actuators & $2.10^{4}$ & -\\
inter-actuator distance & 26 cm & - \\
AO frequency & 2.5 KHz & -\\
wavelength & 1.6 $\mu m$ & 1.6 $\mu m$\\
average Strehl ratio &Ê$83 \% - 96 \%$ & - \\
delay  & 0.04 s  & - \\
focal sampling & 0.125 $\lambda/D$ / pixel & 0.125 $\lambda/D$ / pixel \\
static aberrations upstream of the coronagraph & - &  10 nm rms\\
static aberrations downstream of the coronagraph & - & 0.3 nm rms \\
central obscuration default value & 30\% & 30\% \\
\hline 
\textbf{Studied parameters} & & \\
\hline
central obscuration & 10 - 30 [\%] & 10 - 30 [\%]\\
spider vanes thickness & 30 - 75 [cm] &30 - 75 [cm] \\
segments reflectivity ($\sim$750 of 1.5m diameter)&  1 - 5 [\% ptv] &1 - 5 [\% ptv]\\
segments static aberrations ($\sim$750 of 1.5m diameter)& 6 - 30 [nm rms] &6 - 30 [nm rms]\\
pointing errors & 0.1 - 0.5 [mas rms] &0.1 - 0.5 [mas rms]\\
pupil shear & 0.1 - 0.5 [\%] & 0.1 - 0.5 [\%] \\
\hline
\end{tabular}
\end{table*}
\end{center}

\subsection{Metrics}
In the following, we describe metrics used to evaluate efficiency of coronagraphs. Caution: none of these metrics are weighted by the overall coronagraphic system transmission ($\mathscr{T}$). This throughput is set by the pupil stop transmission (times the mask transmission for BLs). The system transmission (presented in Table. \ref{coro1}) basically remains a physical limitation that must influence the decision for which coronagraph to implement in practice (integration time issue), but here we are more interested on upper limit of coronagraphs for comparison clarity in regards with external limitations. This point will be further discussed in Sec. 
\ref{sec:study2}.
We also apprise the reader that some coronagraph designs may attenuate the planet signal (e.g the four phase transitions between adjacent quadrants of the FQPM create four $\lambda/D$-blind zones or the repetitive throughput-less zones of some particular BLs functions) which is not quantify through our metrics.

\subsubsection{Coronagraphic halo}
\label{metrics:XAO}
Several metrics can be used to quantify the capability of a coronagraph \citep[for instance]{2004EAS....12..165B}. 
At the level of the coronagraphic image we have identified two metrics. The first one, $C_{CORO}(\rho)$ is the contrast profile averaged azimuthally and the second one $\overline{C_{CORO}}$ gives the contrast between the star peak and an average intensity in an annular region of the focal plane where an off-axis companion are expected to be detected. These metrics read : \\

\begin{equation}
C_{CORO}(\rho) = \frac{\int^{2\pi}_{0} \psi_{CORO} (\rho, \alpha) \, d\alpha} { \ 2\pi \, \psi_{PSF} (0)} 
\label{Cr}
\end{equation} 

\begin{equation}
\overline{C_{CORO}} = \frac{\left(\int^{\rho_f}_{\rho_i} \int^{2\pi}_{0} \psi_{CORO} (\rho, \alpha) \rho\,  d\rho\, d\alpha \right) \slash \pi ({\rho_f}^{2} - {\rho_i}^{2})} {\psi_{PSF} (0)} 
\label{C}
\end{equation}

Where $\rho_i$ and $\rho_f$ are the inner and outer radii of the annular region;  $\psi_{PSF}(0)$ is the maximum intensity of the star image on the detector (without the coronagraph, except for the APLC and APRC for which this term includes the apodizer transmission); $\psi_{CORO}(\rho, \alpha)$ is related to the intensity of the coronagraphic image on the detector. 

We use these metrics to study the variation of performance with respect to telescope parameters and as a function of the Strehl ratio. 

The area of calculation in the focal plane for $\overline{C_{CORO}}$ can be well matched to the instrumental parameters (the width of the ring can be modified to match science requirements). For most results presented hereafter, the search area is bounded at $\rho_i = 4 \lambda/D$ for short radii (IWA requirement) and at $\rho_f = 80 \lambda/D$ for large radii (XAO cut-off frequency).
These limits translate to 30mas and 0.63" respectively at $1.6 \mu m$, and allow coronagraphs comparison over a large region of interest while keeping the study general and independent of a specific science requirement. The impact of $\rho_i$ and $\rho_f$ values will be further discussed in Sec. \ref{sec:XAO}.

\subsubsection{Differential Imaging residuals}
\label{metrics:DI}
When using a DI system implying some image subtraction, the average contrast is no longer suited. Results will be presented as radial contrast plots ($5\sigma$ normalized contrast vs. angular separation) to compare coronagraphs:
\begin{equation}
C_{DI}(\rho) = \frac{5 \times \sigma \left[\psi_{CORO_{1}}(\rho) - \psi_{CORO_{2}}(\rho)\right]}{\psi_{PSF}(0)} 
\label{D}
\end{equation}  
Here,  $\sigma[]$ is an operator which denote the azimuthal standard deviation measured in a ring of width $\lambda/D$ on the subtracted image $\psi_{CORO_{1}}-\psi_{CORO_{2}}$. $C_{DI}$ quantifies the ability to pick out an off-axis companion at a given angular distance.  

Here, we adopt that simple metric for sake of clarity but we note that more appropriate criteria adapted to the case of high contrast images have been developed by \citet{2008ApJ...673..647M}.

\subsection{Studied parameters}
\label {paramsstudy}

\subsubsection{Central obscuration}
\label{sec:obscuration}
It is very likely that future high contrast instruments will have to deal with large central obscuration ratios, possibly larger than the current 8-m class telescopes (e.g 30$\%$ for the future E-ELT).
We evaluate its impact on coronagraphic performance for obscuration ranging between 10\% to 30\%.
For each central obscuration ratio, APLC operates at its optimum configuration as defined in \citet{2007A&A...474..671M}. For the range of  obscuration we are considering, the Lyot mask of the APLC varies between 4.3 to 4.9$\lambda/D$. The apodizer of the APRC is also re-optimized for each case. When it is not specified, the default value of the central obscuration ratio is 30\%.

\subsubsection{Spiders thickness}
\label{sec:spiders}
The analysis for the spider thickness is made for one configuration where six symmetrical cables are used to maintain the secondary support \citep[Fig. \ref{OWL}, left]{2004SPIE.5489..391D}. The thickness varies from 30 to 90 cm and for each case and each coronagraph, the pupil stop is re-optimized to match the entrance aperture. In the particular case of BL8, the high order of the mask yields to unusable pupil stops (near to 0$\%$ throughput) to correctly match diffraction when spider vanes are included. Hence, we relax constraints on performance (not anymore perfect at $S = 100\%$) to increase throughput. As a result, performance presented in section \ref{sec:XAO2} and \ref{sec:DI} are affected.

\subsubsection{Segment reflectivity variations}
\label{sec:reflec}
ELT primary mirrors will be necessarily segmented and amplitude variations are expected due to a difference of reflectivity between the segments (optical coating). 
The variation of reflectivity through an optical system induces wavefront amplitude variation that lead to potentially bright static speckles in the focal plane of the instrument. 
It is important to know how robust a coronagraph is to these defects. 

\begin{figure}[!ht]
\begin{center}
\includegraphics[width=4cm]{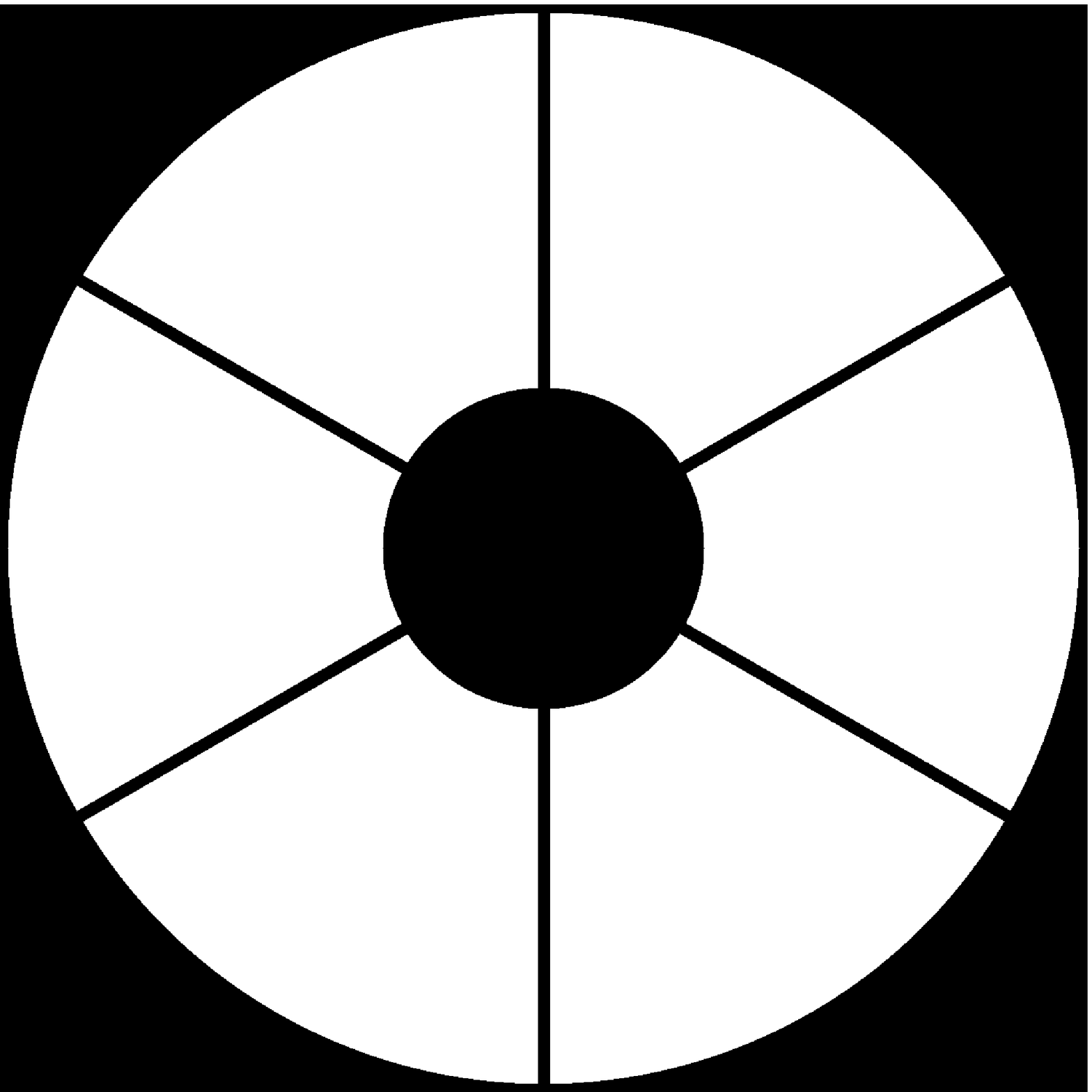}
\includegraphics[width=4cm]{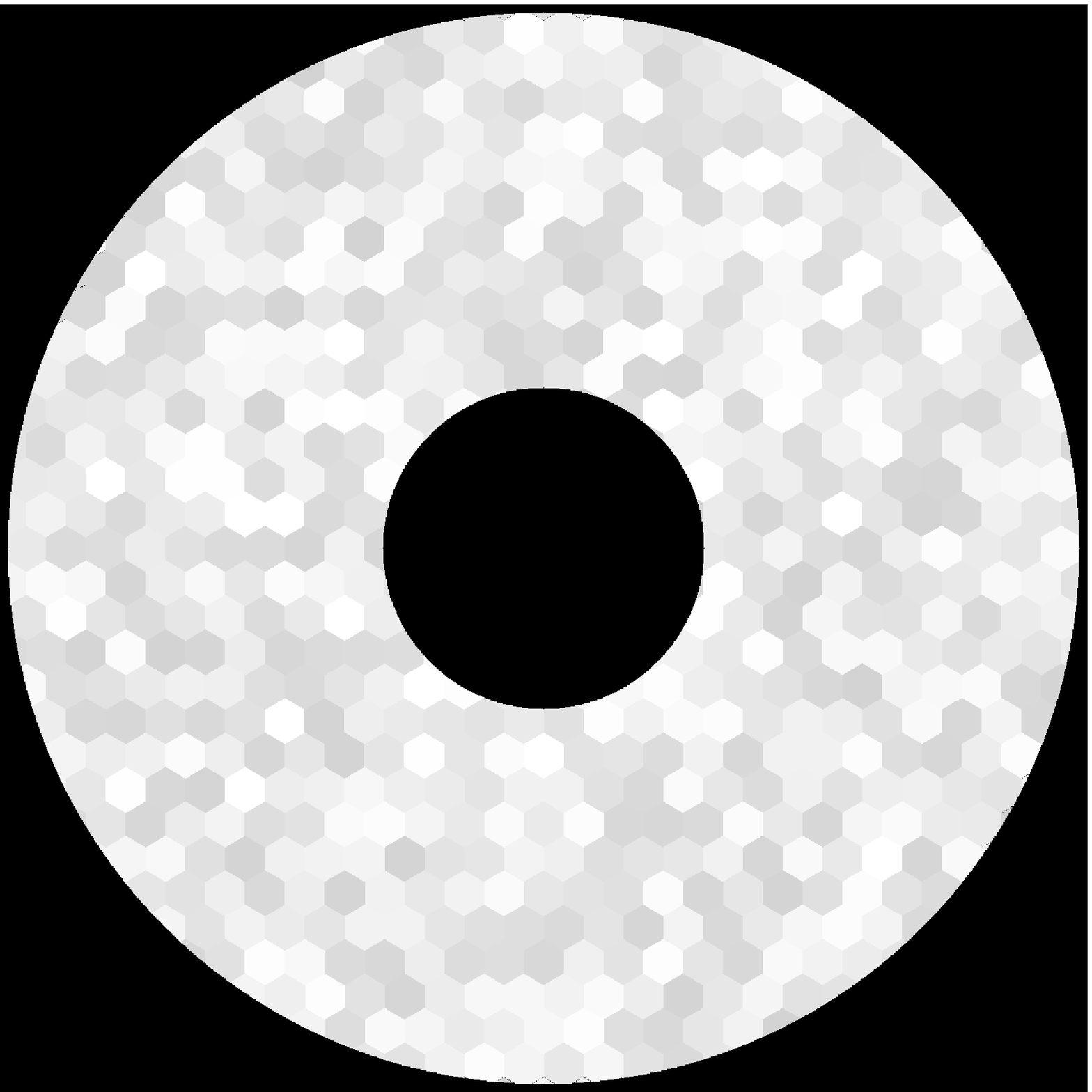}
\end{center}
\caption{Spider vanes configuration (left) considered in Sec. \ref{sec:spiders}, pupil with reflectivity variations (right) considered in Sec. \ref{sec:reflec} (levels are exagerated for sake of clarity). } 
\label{OWL}
\end{figure}

We assume $\sim750$ hexagonal segments of 1.5 meters diameter (Fig. \ref{OWL}, right) and assess the impact of a uniform 
segment-to-segment reflectivity variation from 1\% to 5\% (peak-to-valley, hereafter ptv). For comparison,  5\% (ptv) is the typical variation measured on the Keck telescope \citep{2003SPIE.4840...81T}.

\subsubsection{Segment static aberrations}
\label{sec:aberrations}

Segment aberrations refer to low-order static aberrations (piston, tip-tilt, defocus and astigmatism) producing speckles that fall relatively near the central core of the image. 
Higher order aberrations are not considered at this stage but will be implemented in the DI simulations.
The limited number of actuators in the AO system imposes a control radius in the image plane which scales as $N/2$ if $N$ is the linear number of actuators across the pupil. 
We assume that the static aberrations are already well corrected inside this radius. To estimate the actual segment aberrations corrected by the XAO system, we set the PSD of the phase to a null contribution at frequencies lower than the cut off frequency. This method gives the best possible correction that can be obtained (only limited by the fitting error and it does not include any wavefront sensor errors, influence function nor actuators space positioning).
An analytic study of AO correction of segment errors was performed in \citet{2006SPIE.6267E..86Y} where the quantity $\mathscr{\gamma}_{AO}$, the amount of wavefront correction achieved by an AO system on a particular segment aberration, was defined by the ratio of the corrected rms wavefront error to the initial uncorrected rms wavefront error:
\begin{equation}
\mathscr{\gamma}_{AO} = \frac{\sigma_{corrected}}{\sigma_{initial}}
\label{AOequ}
\end{equation}

\begin{figure}[!ht]
\begin{center}
\includegraphics[width=4cm]{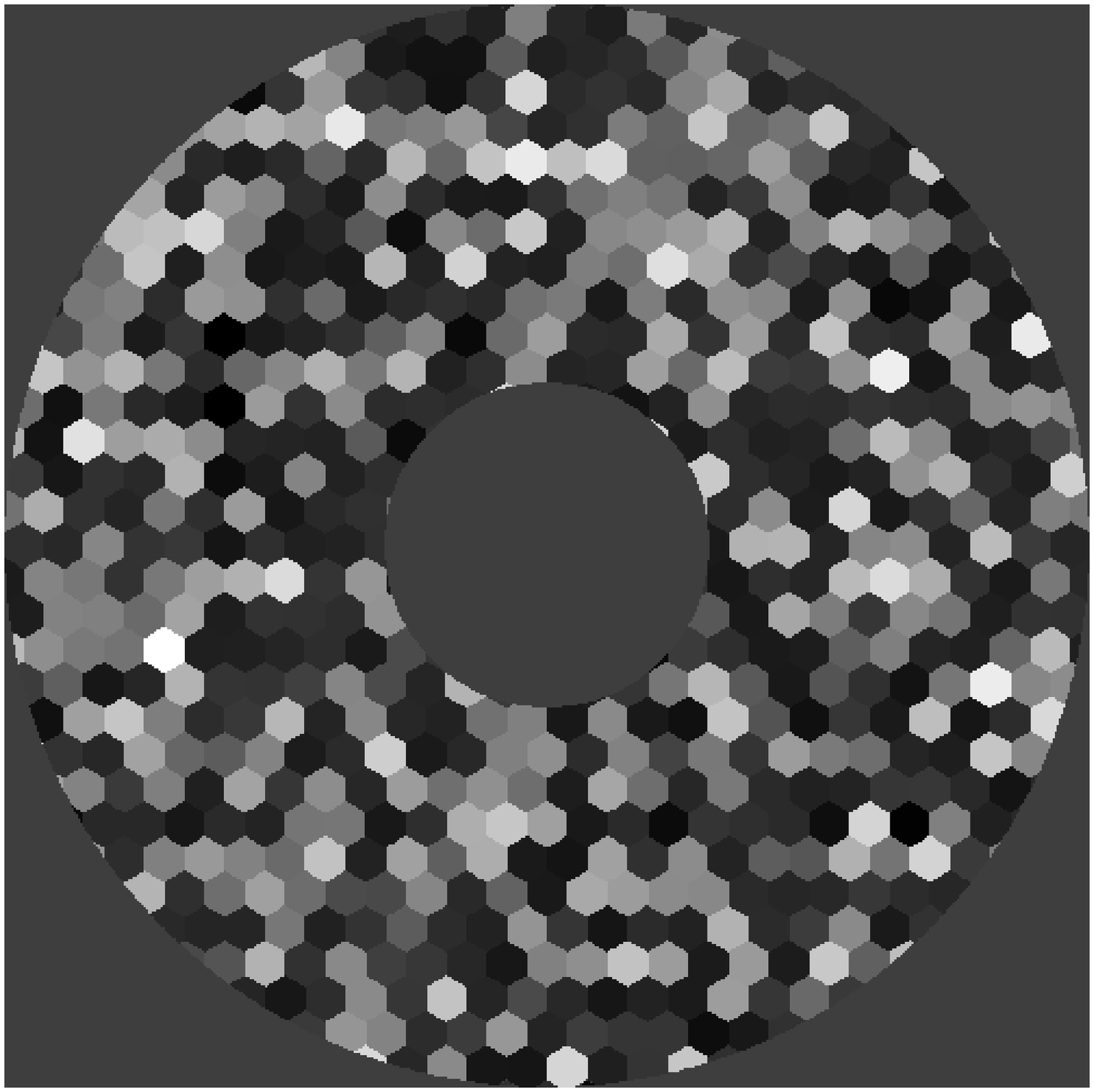}
\includegraphics[width=4cm]{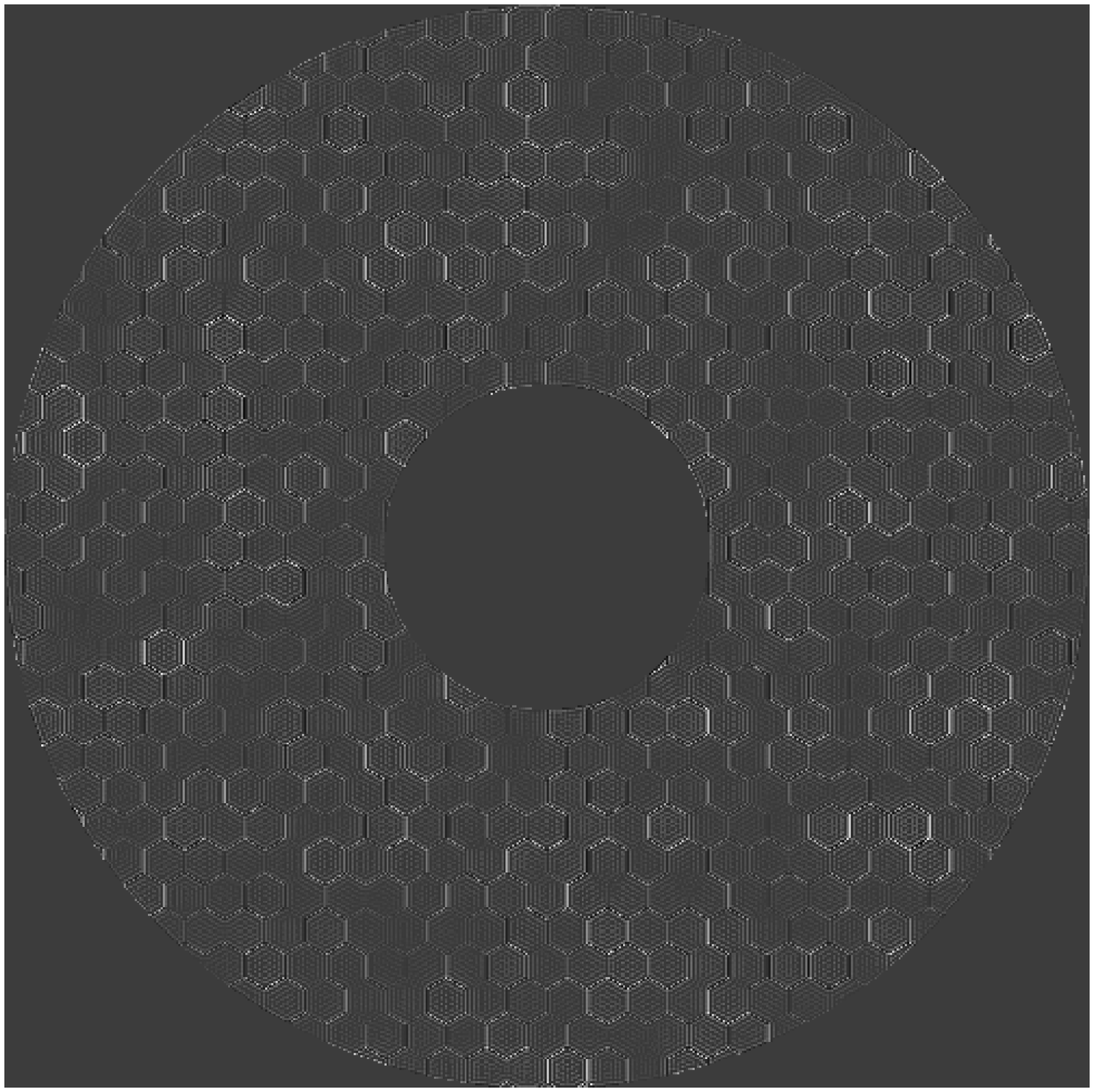}
\end{center}
\caption{left: Initial piston WFE of 30 nm rms, right: Residual piston of 6.7 nm rms after XAO correction (levels are exagerated for sake of clarity).} 
\label{OWL2}
\end{figure}

With the XAO system we have considered, the analytical method yields a theoretical value of $\sim0.32$ for piston-like phasing error. With our simple "PSD shaping" we obtained the following values for $\mathscr{\gamma}_{AO}$: 0.22, 0.34, 0.27, 0.41 for piston, tip-tilt, defocus and astigmatism respectively which is not that much different to the analytical value. 
As a result, the XAO system significantly reduces the wavefront error of each segment as show in Fig. \ref{OWL2} (piston phasing errors example). 

Predicting the level of low-order aberrations that ELT segments will feature is quite difficult. Nevertheless, measurements with Keck telescope \citep{2000SPIE.4003..188C} show that 10 nm rms is reachable. In our simulations, we consider a range of intial wavefront error from 6 to 30 nm rms, which is corrected by the XAO system and hence reduced to values ranging from 0.7 to 12 nm rms. In practice, we study each static aberration independently from each other, and find an undistinguishable impact on coronagraphic halo, thus only the case of piston will be presented.

\subsubsection{Pointing errors and finite size of the star}
The offset pointing error refers to the misalignment of the optical axis of the coronagraph with the star (here, we assumed that the star is a point source like). For instance, with SPHERE the goal is 0.5 mas rms, hence a direct translation of this requirement to a 42 meters telescope, would be a pointing error residual of less than 0.1 mas rms. In practice, we evaluate the effect of pointing error between 0.1 and 0.5 mas rms.

If the star is not point-like but its disk is slightly resolved it can be modeled as a sum of incoherent off-axis point sources.
As for the offset pointing error, coronagraphs that allow a very small IWA will be more affected. Actually, the impact of the finite size of the star is quite similar to the one of the offset pointing.

\subsubsection{Pupil shear}
\label{stop}
Most of coronagraphs include several optical components: apodizer, focal plane mask and pupil stop.
As a result their performance also depends on the alignment of these components. 
\noindent In particular, the pupil stop has to accurately match the telescope pupil image. This condition is not always satisfied, and the telescope pupil may undergo significant mismatch which could amount to more than a  few \% of its diameter. 
The pupil shear is the mis-alignment of the pupil stop with respect to the telescope pupil image. It is an issue especially for ELTs for which mechanical constraints are important for the design. For example, the James Webb Space Telescope is expected to deliver a pupil image for which the position is known at about 3-4\%. Therefore, the performance of the mid-IR coronagraph \citep{2004EAS....12..195B} will be strongly affected. On SPHERE, the planet-finder instrument for the VLT (2010), the pupil shear was identified as a major issue and a dedicated Tip-Tilt mirror was included in the design to preserve the alignment at a level of 0.2\% \citep{2006tafp.conf..353B}. In this study we consider a range of misalignment between 0.1 and 0.5\% of the pupil diameter.


\section{Results}
\label{sec:study1}

\subsection{XAO simulation}
\begin{figure*}[!ht]
\includegraphics[width=9cm]{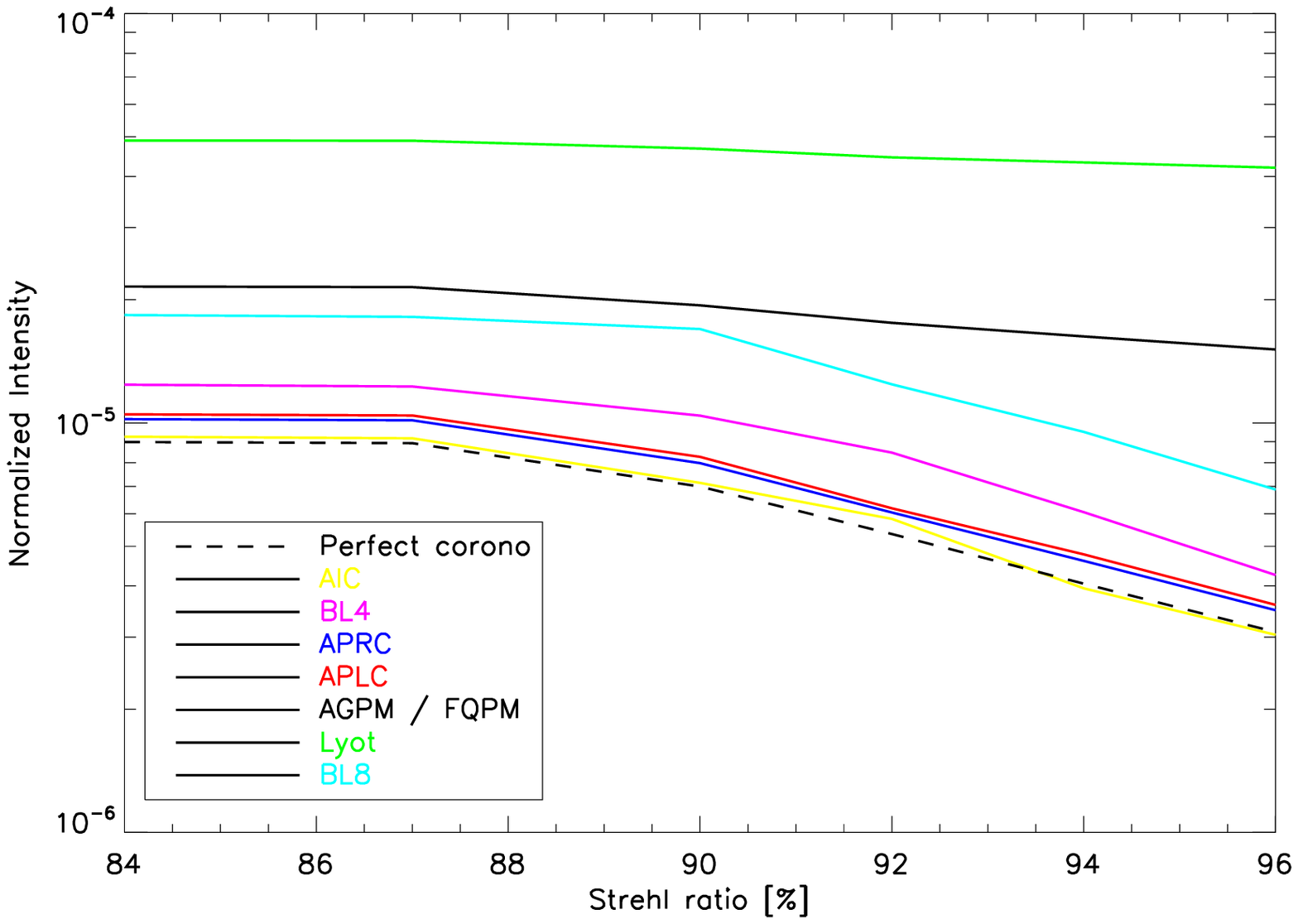}
\includegraphics[width=9cm]{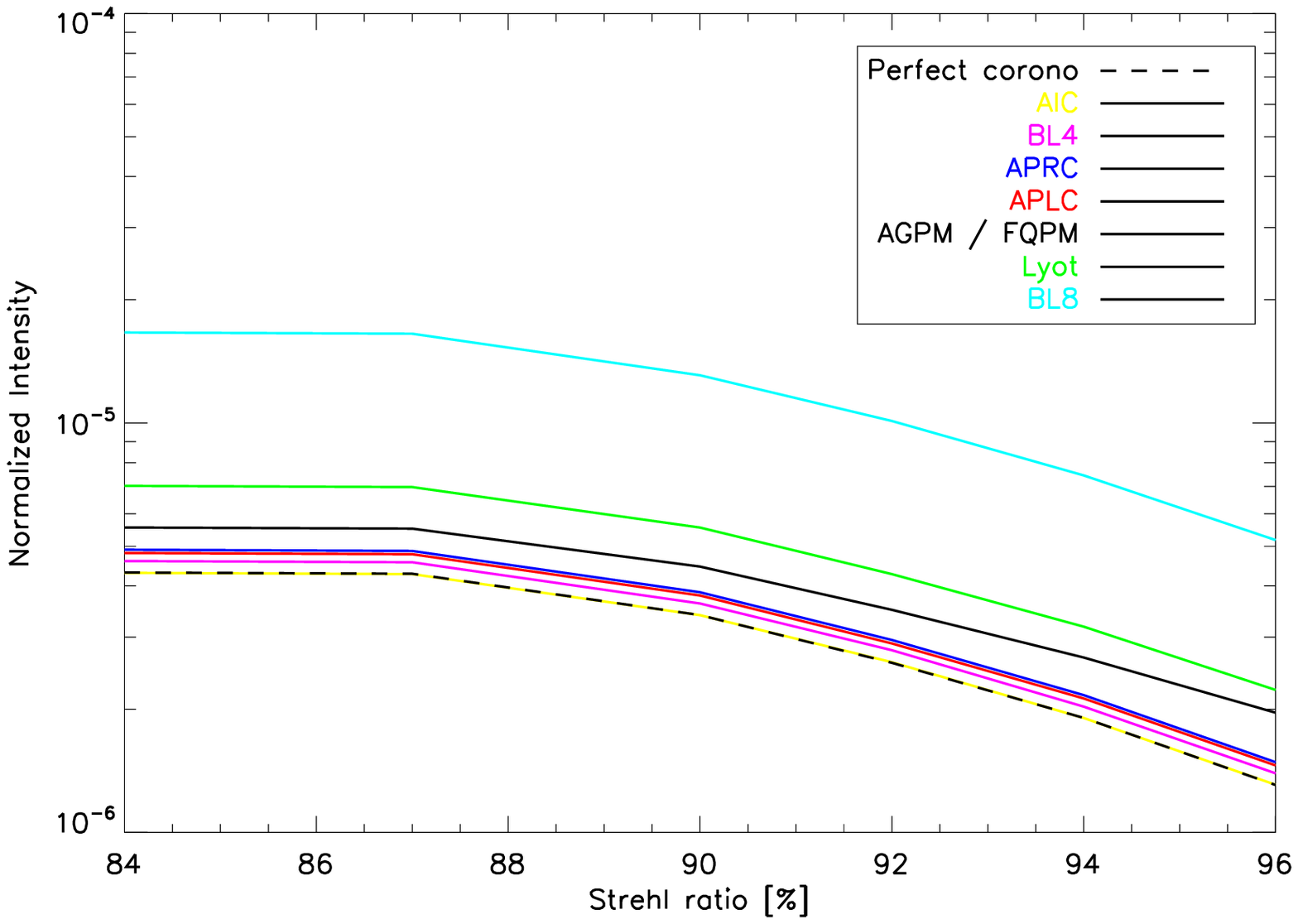}
\caption{Variation of $\overline{C_{CORO}}$ as a function of the Strehl ratio for all coronagraph concepts. Left: at 4$\lambda/D$ (IWA), Right: averaged from 4$\lambda/D$ (IWA) to 80$\lambda/D$ (AO cut-off frequency).} 
\label{Perf1}
\end{figure*}

\subsubsection{Influence of the wavefront correction quality}
\label{sec:XAO}
We first started to compare the coronagraphic performance as a function of the Strehl ratio ($S$) with the $\overline{C_{CORO}}$ metric. 
The objective of this first analysis is to assess the raw contrast delivered for each coronagraph considering only the diffraction by the edges of the pupil and the residual atmospheric phase aberrations which is leaking through the XAO system. Therefore, these defects will produce a perfectly averaged halo of speckles which sets the level of the photon noise in the coronagraphic image plane. Obviously the contrast level must be much better than this coronagraphic halo but this noise contribution estimate will be necessary to investigate the signal to noise ratio achievable for detecting exoplanets with ELTs in further studies. 

Figure  3 shows $\overline{C_{CORO}}$ as a function of the Strehl ratio for two locations in the coronagraphic image. At the left, for an angular separation of 4$\lambda/D$=IWA and to the right, averaged in between the IWA and the AO cut off frequency ($\mathscr{F}_{AO}=80\lambda/D$). In each case, $\overline{C_{CORO}}$ for a perfect coronagraph is plotted in dashed line. This ideal model is helpful since it reveals the limitations from the residual aberrations that are leaking through the XAO system
only (i.e by principle there is no pupil edges diffraction contribution since all the coherent part of the light has been removed). The actual contribution on the limitations sets by the diffraction of the edges of the pupil is actually revealed by the discrepancy of real coronagraph $\overline{C_{CORO}}$ curves to that of the ideal model.
Two regimes can be identified:\\
1/ where $\overline{C_{CORO}}$ of a coronagraph follows $\overline{C_{CORO}}$ of the ideal model, which corresponds to the speckle dominated regime where coronagraphs perform much better than the XAO and so the performance is set by the XAO itself. In other words, improving the XAO correction is necessary to improve final performance. In such a case, contrast increases with $S$ and a substantial gain in starlight suppression imposes to reach a high level of wavefront correction ($S\sim$94\%). For the considered range of $S$ values (84 - 96$\%$) all the coronagraphs considered are in this regime (Fig. \ref{Perf1}, right) except for the AGPM/FQPM and the Lyot but only when $\overline{C_{CORO}}$ is evaluated at the IWA (Fig. \ref{Perf1}, left).\\
2/ the diffraction dominated regime appears when $\overline{C_{CORO}}$ of a coronagraph does not anymore follow $\overline{C_{CORO}}$ of the ideal model and is about flat, i.e phase aberrations are small enough to reveal the actual limitation of the coronagraphs and so the limitation is mostly set by the diffraction by the edges of the pupil. In other words, improving the XAO correction is useless since the limitation comes from the coronagraph (AGPM/FQPM and Lyot cases previously underlined at IWA).

The particular case where $\overline{C_{CORO}}$ is about flat while still following the ideal model (below $S=88\%$) corresponds to a case where the limitation comes from the residual phase aberrations that are present in a so large amount that improving the XAO correction (from 84 to 88$\%$) does not yield to an improvement of the performance.

The AGPM/ FQPM and Lyot coronagraphs have a strong dependency with the area where $\overline{C_{CORO}}$ is evaluated which indicates that most of the residual energy is actually localized near the image center in contrary to other coronagraphs. This is a consequence of the diffraction by the central obscuration which is not favorable to such designs. At angular distances larger than 4$\lambda/D$, the AGPM/FQPM and the Lyot perform as well as other designs. Thus, the choice performed in Sec. \ref{metrics:XAO} for the value of $\rho_i$ and $\rho_f$ allows a more homogeneous comparisons of coronagraphs.

The contrast achieved with the BL8 is significantly lower than with other coronagraphs. To operate with a 30\% central obscuration and a somewhat small IWA of $4\lambda/D$ the BL8 requires a very aggressive pupil stop ($\mathscr{T}=13.8$). Although, this optimization provides a very deep contrast in a perfect situation when phase aberrations are negligible (S = 100\%) it is no longer the case in a realistic condition, even at high Strehl ratios. This is obviously true for any concepts but the decrease of contrast between the perfect and realistic situations is even more abrupt with the BL8.

\begin{figure*}[!ht]
\centering
\includegraphics[width=7.65cm]{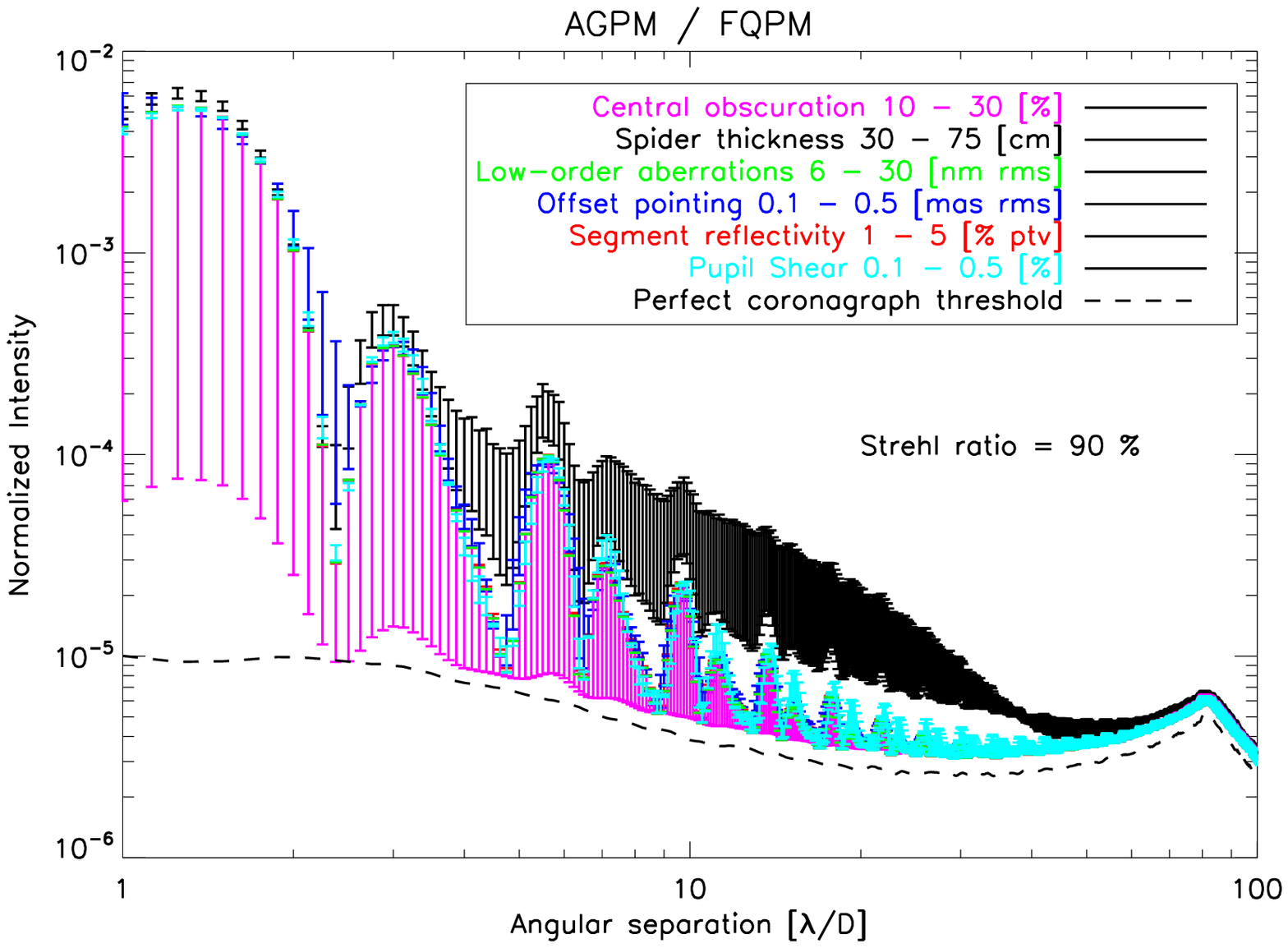}
\includegraphics[width=7.65cm]{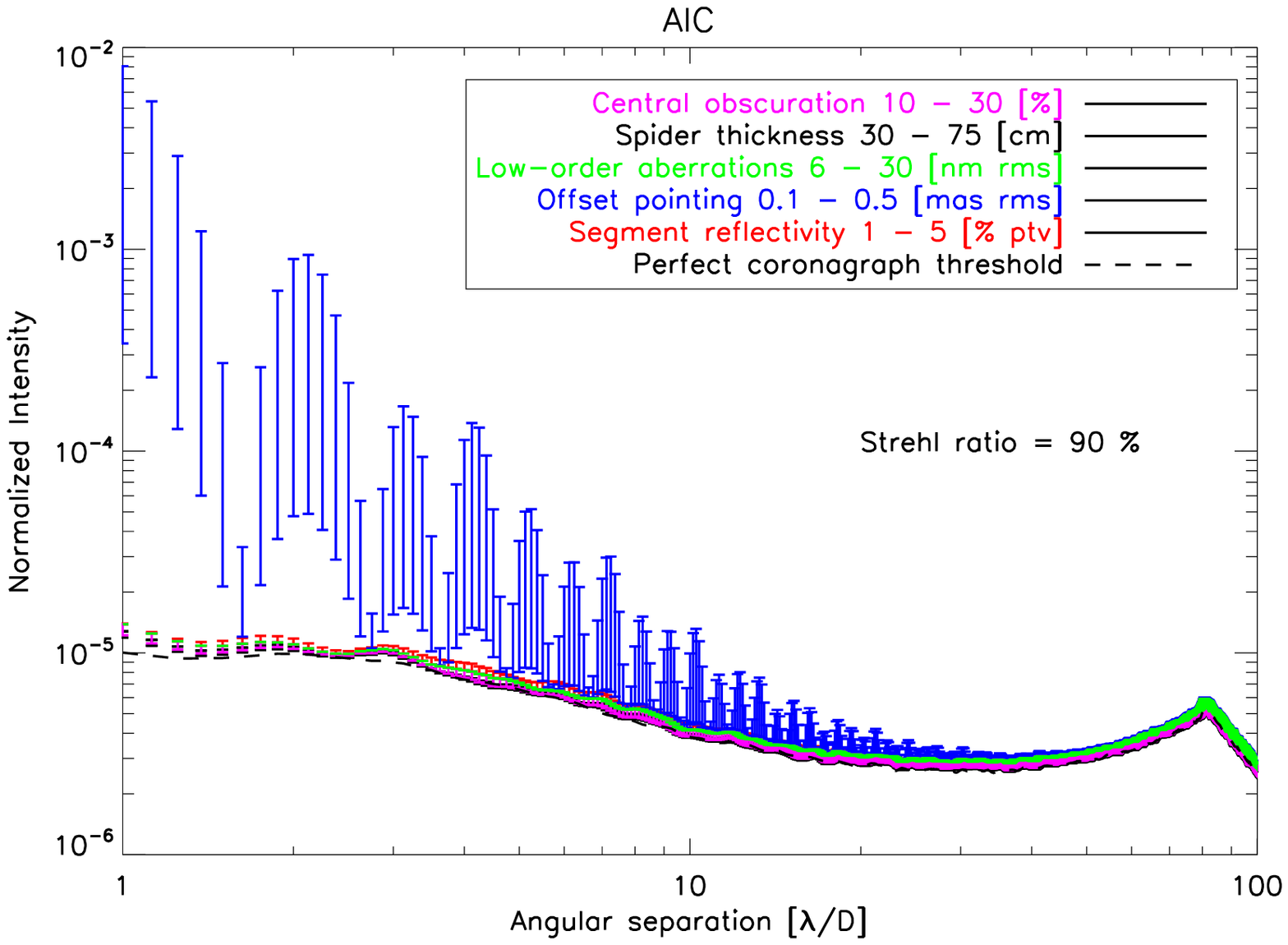}
\end{figure*} 
\begin{figure*}[!ht]
\centering
\includegraphics[width=7.65cm]{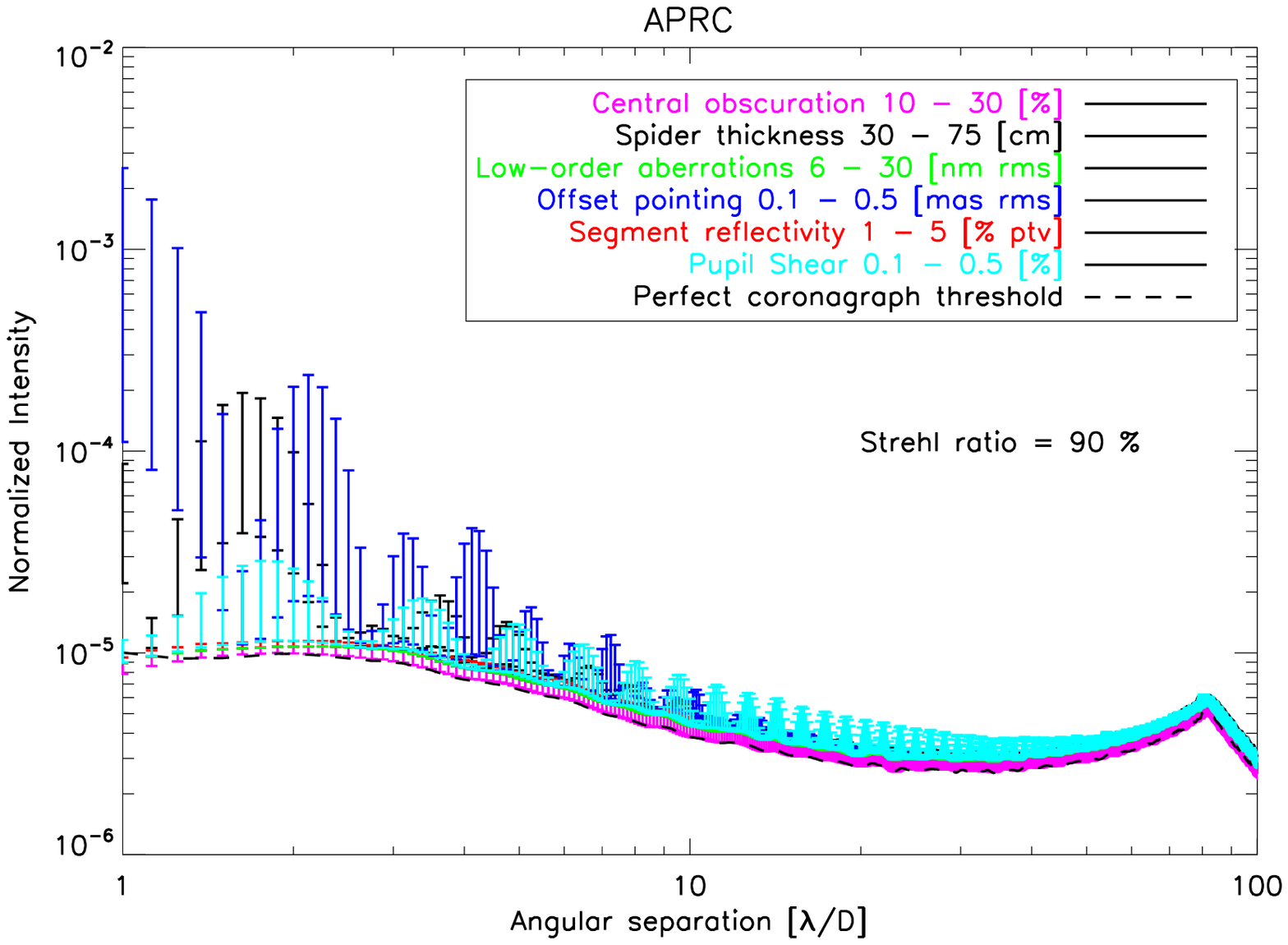}
\includegraphics[width=7.65cm]{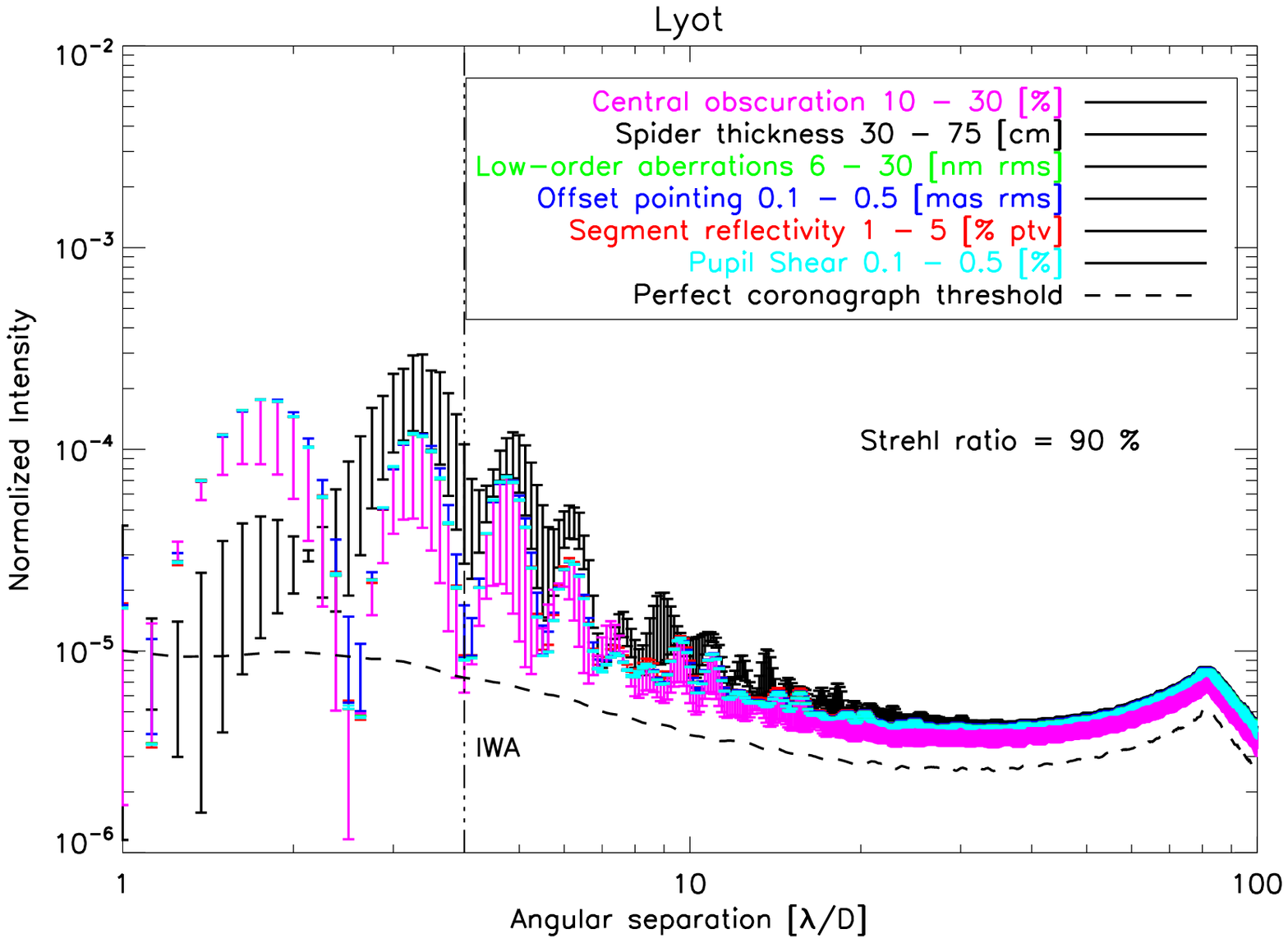}
\end{figure*} 
\begin{figure*}[!ht]
\centering
\includegraphics[width=7.65cm]{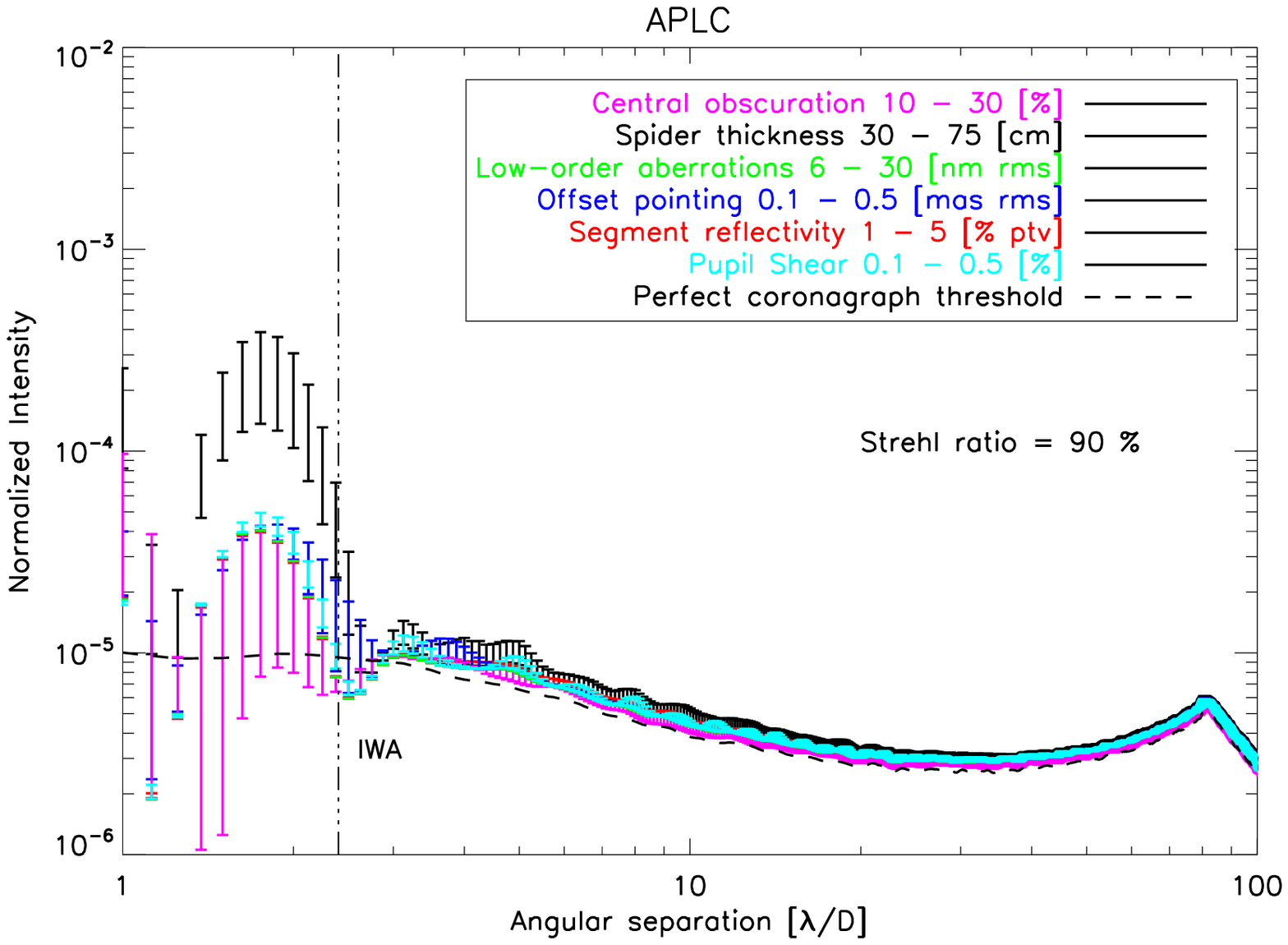}
\includegraphics[width=7.65cm]{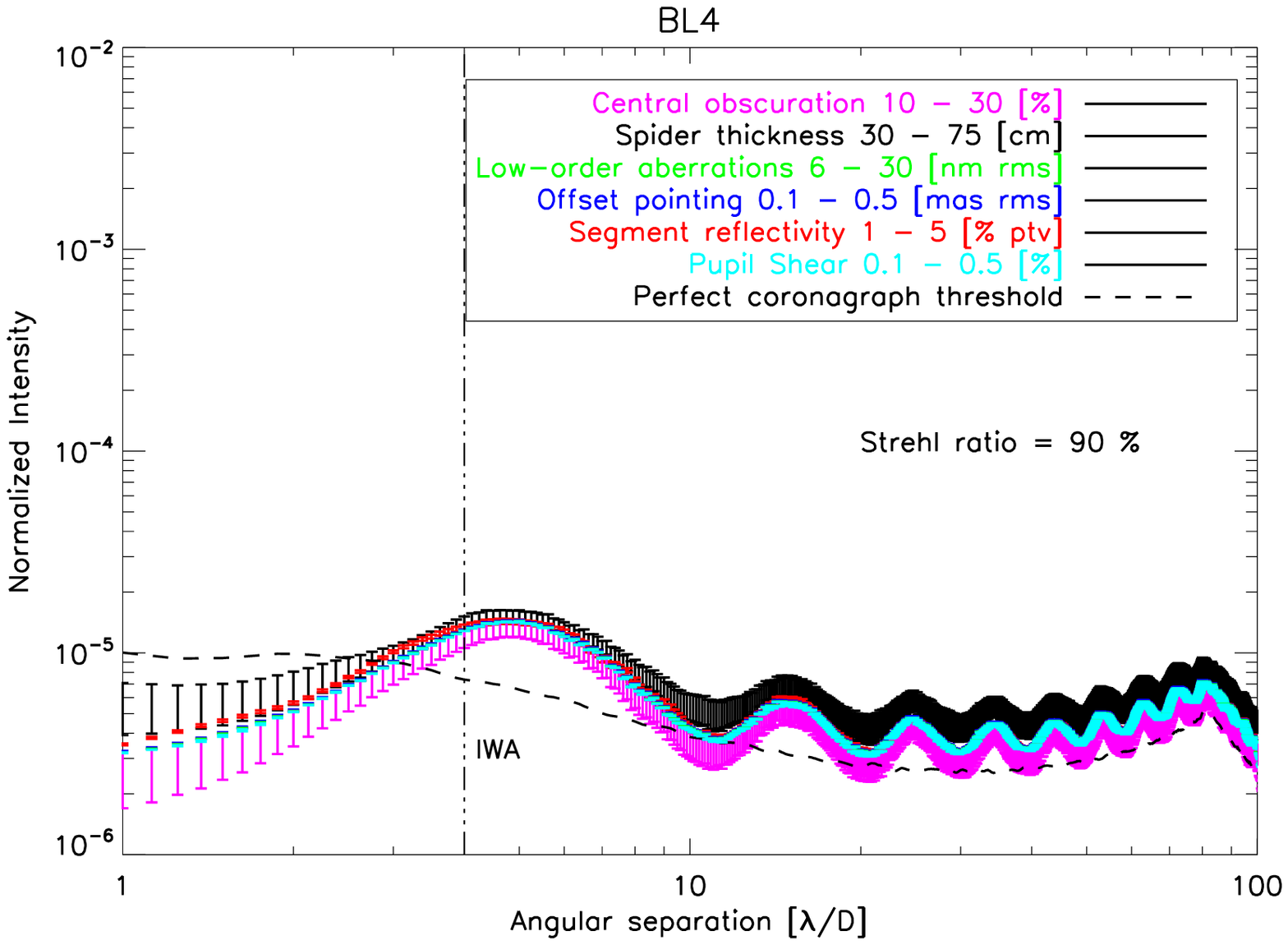}
\end{figure*} 
\begin{figure*}[!ht]
\centering
\includegraphics[width=7.65cm]{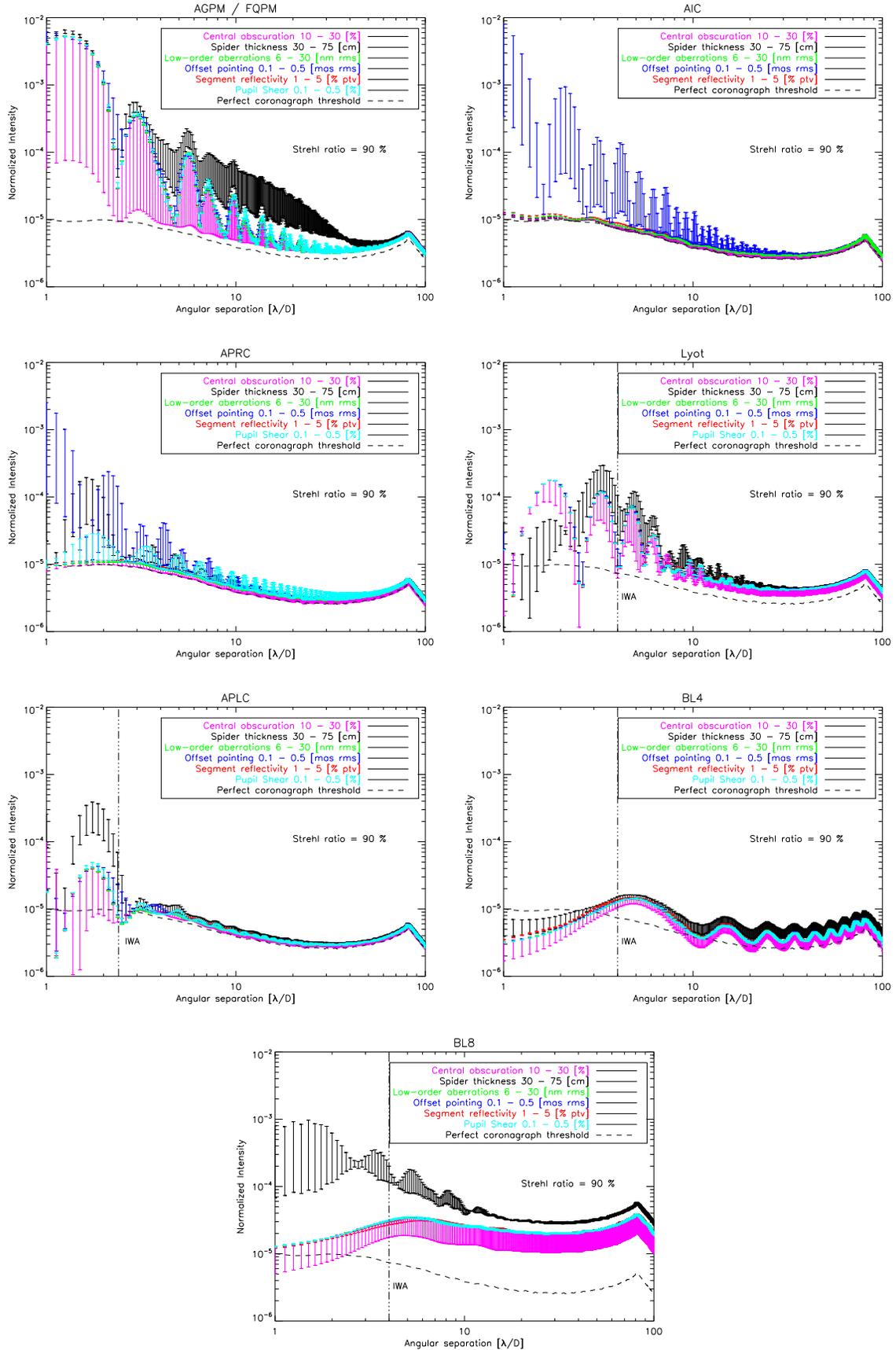}
\caption{Contrast profile $C_{CORO}(\rho)$ (right) for the different parameters and for the following coronagraphs: AGPM/FQPM, AIC, APRC, Lyot, APLC, BL4 and BL8. The Strehl ratio is 90 \%. Error bars indicates the amplitude of the contrast variation. The dashed line stands for the ideal case. } 
\label{coroXAO}
\end{figure*}


\subsubsection{Parameter dependencies}
\label{sec:XAO2}
Figure \ref{coroXAO} analyzes in detail the impact of each parameters defined in section \ref{paramsstudy} for the particular case of S=90\% using $C_{CORO}(\rho)$ the coronagraphic contrast as a function of the angular separation. The variation of each parameter is represented with error bars indicating the dispersion of contrast. Therefore, the sensitivity of $C_{CORO}(\rho)$ to each parameter within a given range of amplitude is shown as an error bar. The range is given
in the legend and is identical for each coronagraph.

For each case, the limit of detection achievable with a perfect coronagraph is plotted as a dashed line.

The specific case of $\overline{C_{CORO}}$ which shows the same quantity as in Fig. \ref{Perf1} but for the several parameters of section \ref{paramsstudy} independently is not plotted to simplify the paper. 
The variation of contrast as a function of the Strehl ratio is actually identical for all parameters and all coronagraphs. In other words, the curves are parallel in between each others and parallel to that of Fig. \ref{Perf1}. This simply means that the contrast is most of the time dominated by the XAO halo and the diffraction by the edges of the pupil. For all coronagraphs but AIC, APRC and APLC, the parameters impacts the contrast at various levels.

In the following, we describe the impact for each coronagraph in the case of S=90\% (Fig.  \ref{coroXAO}):
\smallskip

\noindent \textit{AGPM / FQPM} -- 
At distances shorter than 3$\lambda/D$, the image is dominated by the diffraction of the central obscuration while beyond the contrast limit is set by the spider diffraction spikes. For worst values of dominating parameter the contrast reaches  $2.10^{-4}$ at $4\lambda/D$ and only improves by a factor of 2 at 10$\lambda/D$.
Achievable contrast is quite far from the ideal model (dashed line). 
\smallskip

\noindent \textit{AIC} -- In that case, pupil diffraction is negligible as far as it is centro-symmetrical, but pointing errors are clearly dominating the contrast up to  20$\lambda/D$. The impact is as large as that of the central obscuration for the AGPM/FQPM. At larger distances, the performance of the AIC is identical to that of a perfect coronagraph. 
At $4\lambda/D$ the contrast is $7.10^{-5}$ while at 10$\lambda/D$ it is improved to $2.10^{-5}$. 
\smallskip

\noindent \textit{APRC} -- 10$\lambda/D$ sets the limit between pointing error dominated regime  and pupil shear dominated regime. At $4\lambda/D$ the contrast reaches $4.10^{-5}$ and $9.10^{-6}$ at 10$\lambda/D$. 
 \smallskip

\noindent \textit{Lyot} -- Spider diffraction limits the contrast at any angular radius. However, the impact is not that much important. 
The central obscuration has also a significant signature. At $4\lambda/D$ (near the mask edges) the contrast is only $1.10^{-4}$ but it improves by a factor of 10 at 10$\lambda/D$. 
\smallskip

\noindent \textit{APLC} -- The dispersion of contrast is negligible in that case for any parameter. The APLC achieves the same performance as the perfect coronagraph beyond the IWA and does not feature a dominant parameter. At $4\lambda/D$ the contrast reaches $\sim 1.10^{-5}$ and $5.10^{-6}$ at 10$\lambda/D$. 
\smallskip

\noindent \textit{BL4} -- As for the APLC, the contrast is very close to the perfect case and the dispersion of contrast is small with however a dominance of the spider diffraction spikes. 
At $4\lambda/D$ the contrast is $2.10^{-5}$ while at 10$\lambda/D$ is improved to $6.10^{-6}$.
\smallskip

\noindent \textit{BL8} --The spider diffraction dominates significantly at any angular separations.  The contrast is much worse than for the BL4 and reaches $2.10^{-4}$ at 4$\lambda/D$  and $5.10^{-5}$ at 10$\lambda/D$. 
The BL8 suffers from a severe reduction of the pupil stop therefore distorting the off-axis PSFs while reducing the throughput. High order BLs are actually not suited for ELTs. 
\smallskip

The impact of spider diffraction must be mitigated since contrast profile are azimuthally averaged and therefore 
some image areas features larger contrasts.
Planets could be observed within the clear areas between the spider spike diffractions. This choice depends on the observing strategy.
   
For all coronagraphs, amplitude and phase aberrations on segments in the considered range have much less impact than the diffraction by the pupil edges (central obscuration and spiders). 
For the small IWA coronagraphs, the pointing error is the most dominant factor.

\begin{figure*}[!ht]
\centering
\includegraphics[width=7.65cm]{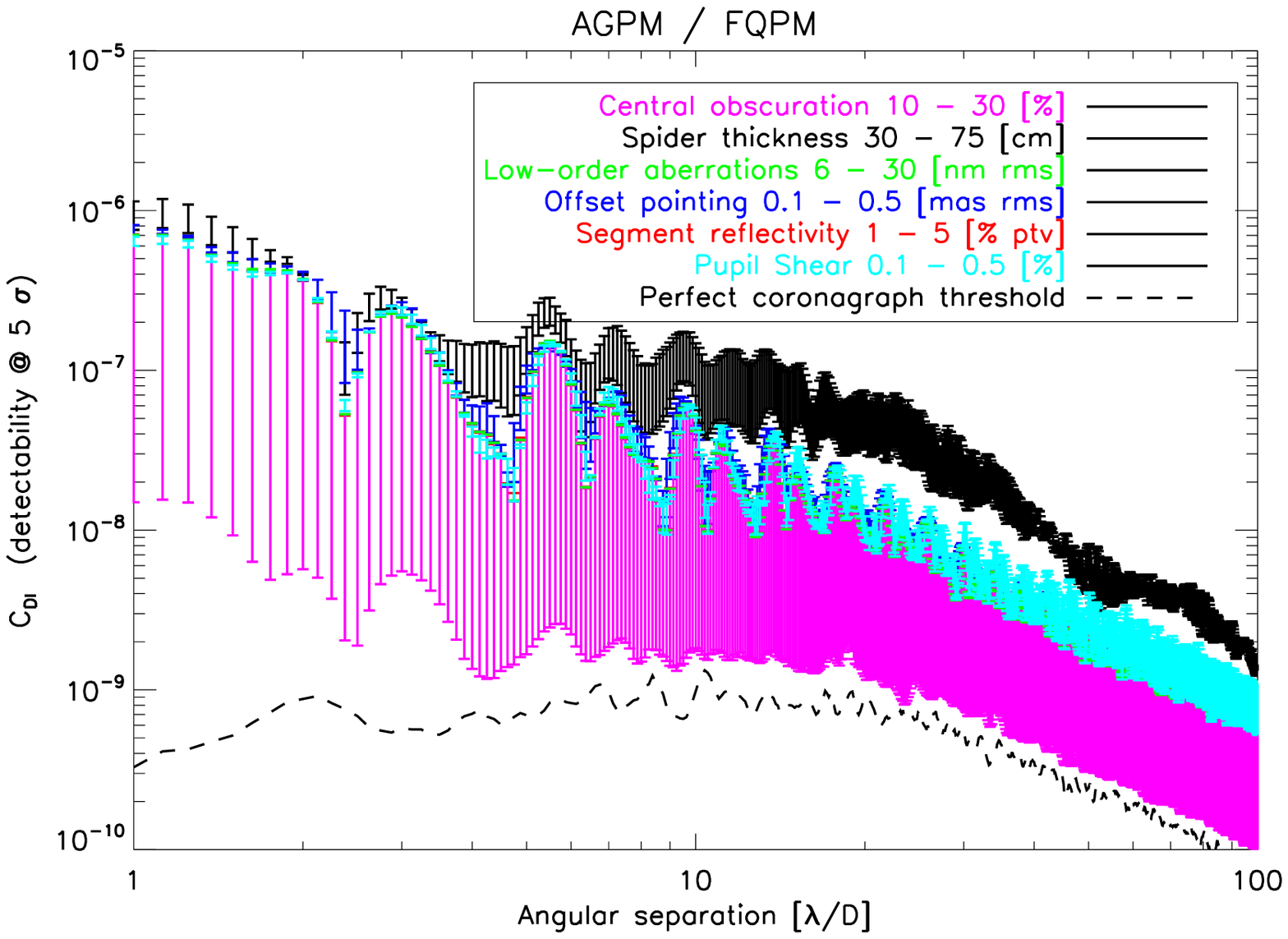}
\includegraphics[width=7.65cm]{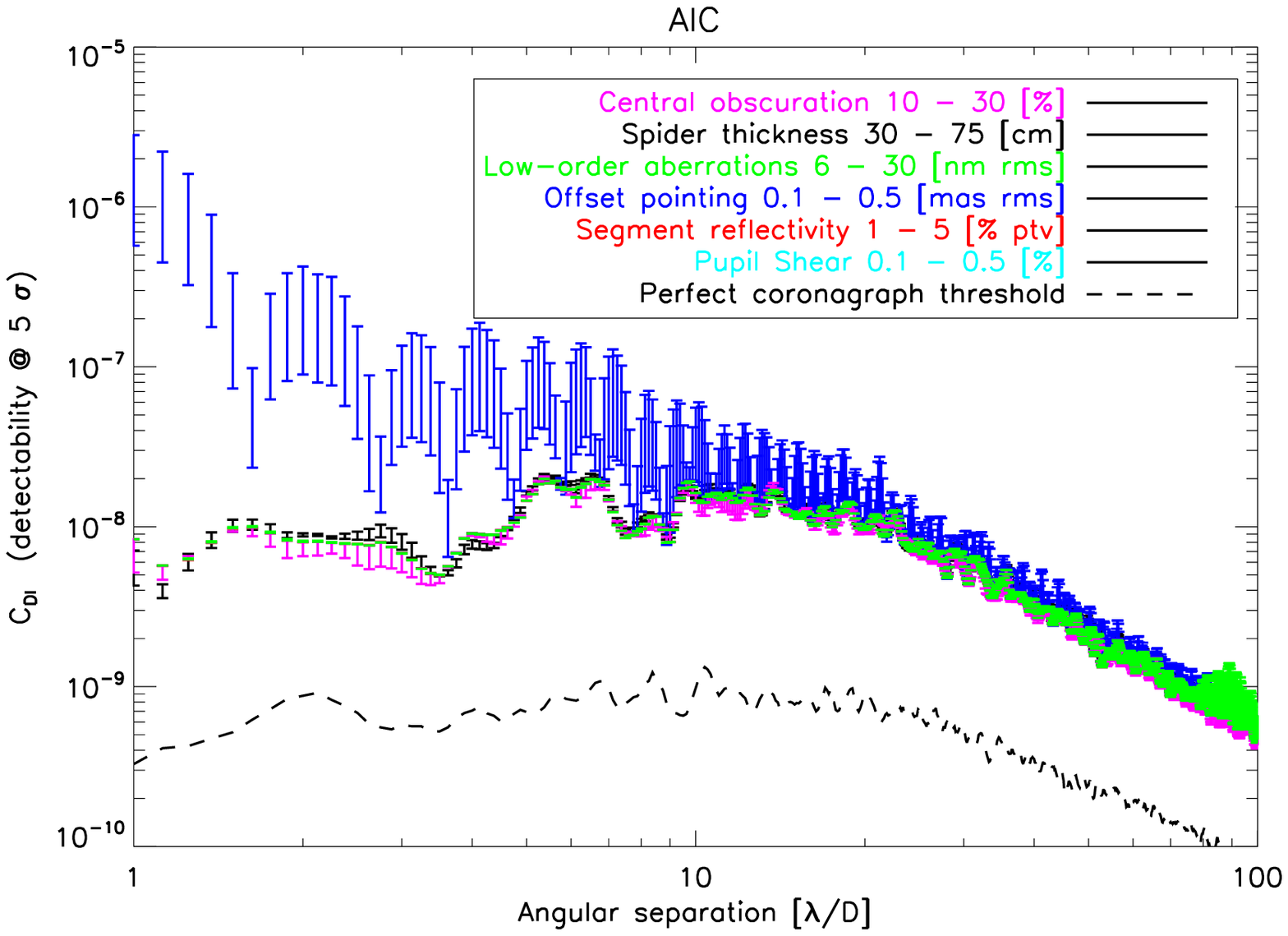}
\end{figure*} 
\begin{figure*}[!ht]
\centering
\includegraphics[width=7.65cm]{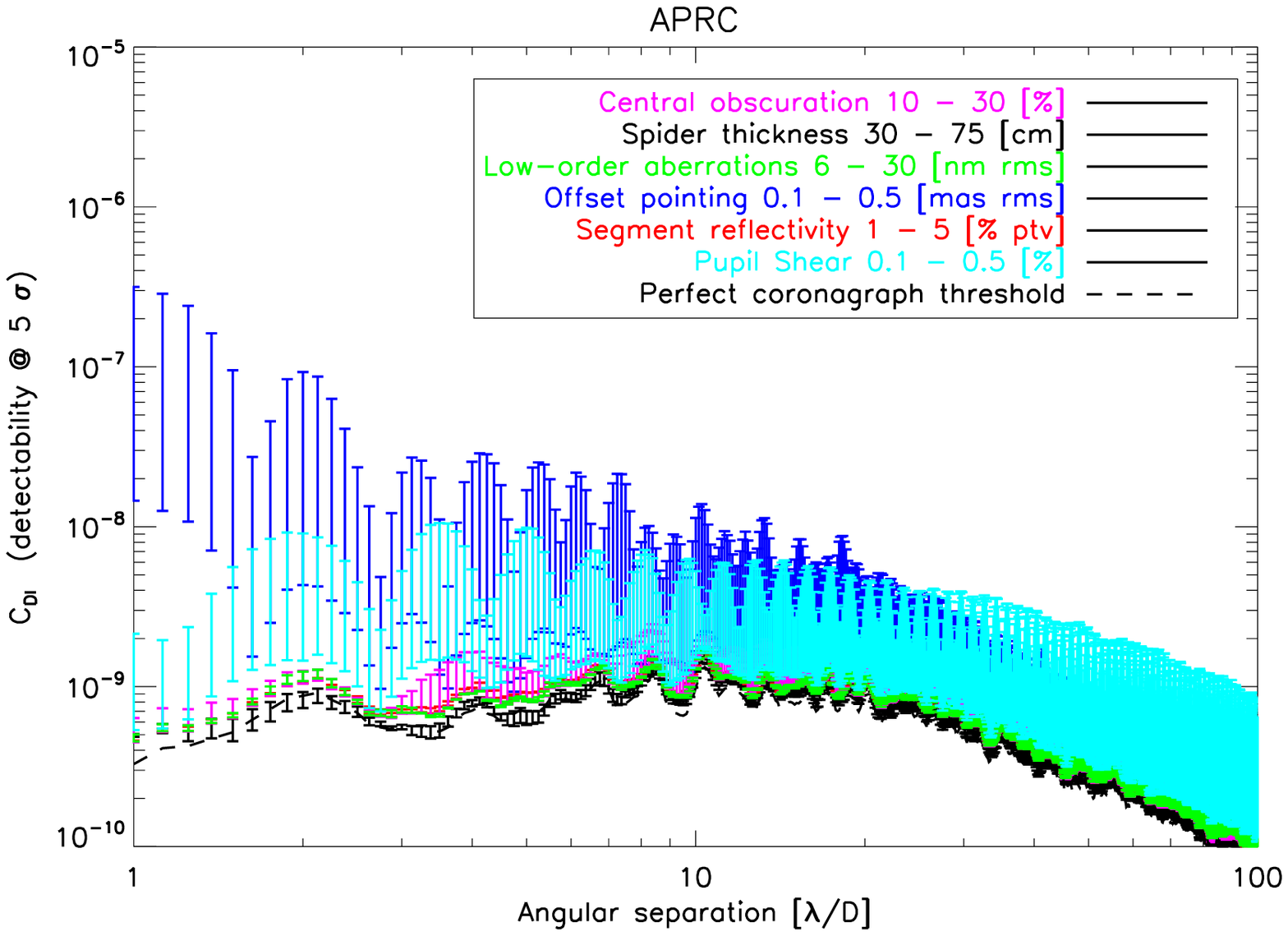}
\includegraphics[width=7.65cm]{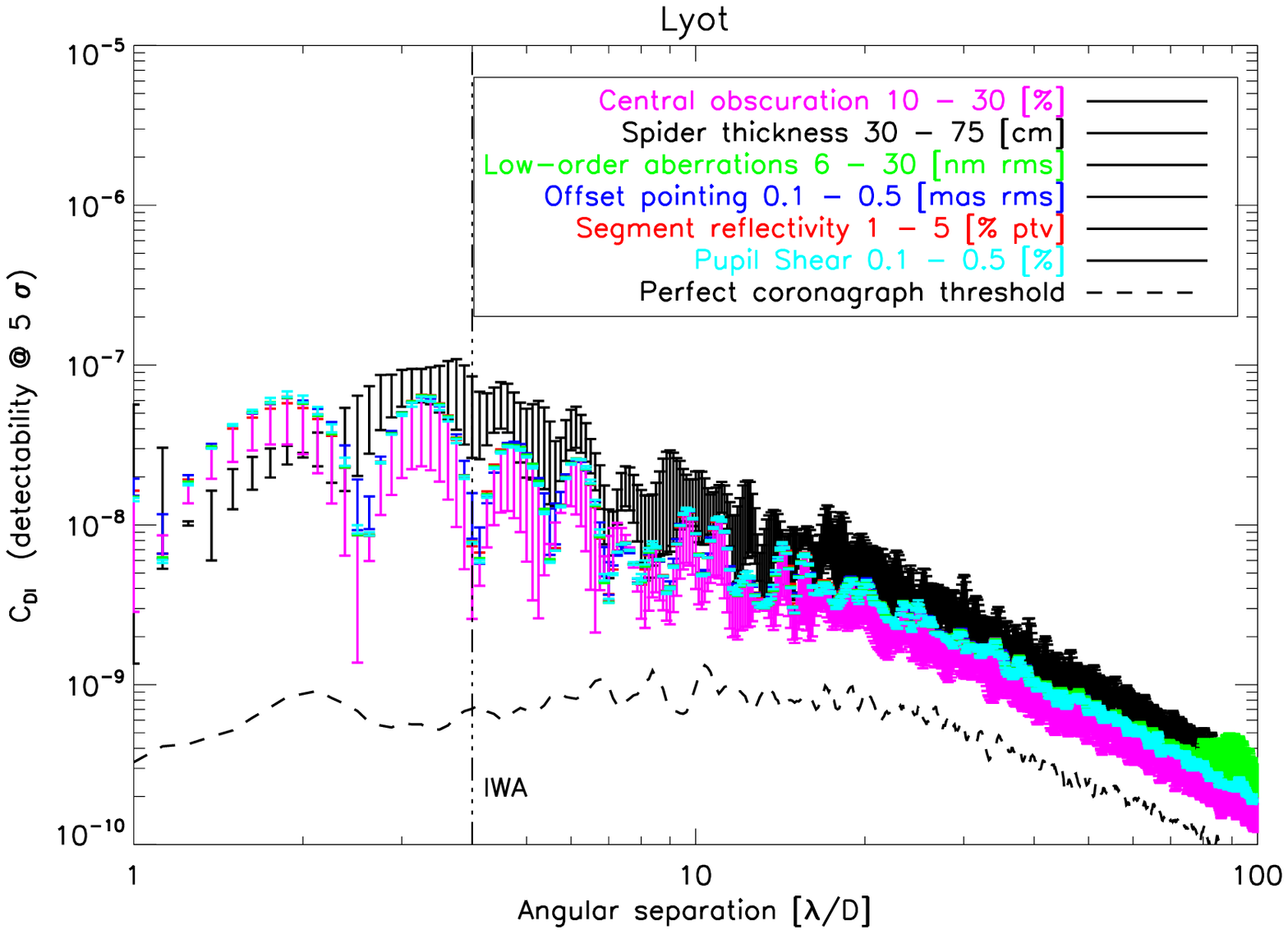}
\end{figure*} 
\begin{figure*}[!ht]
\centering
\includegraphics[width=7.65cm]{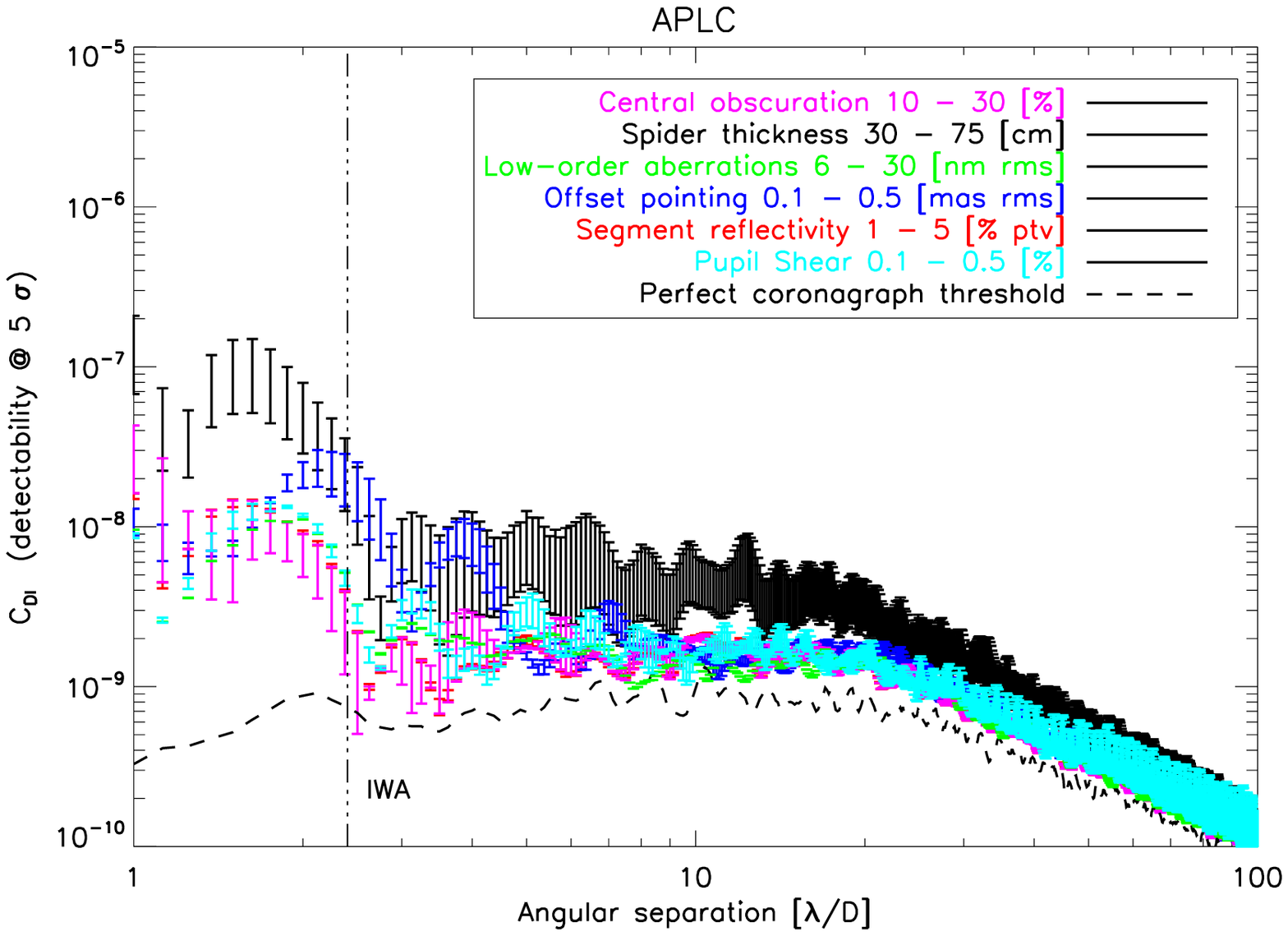}
\includegraphics[width=7.65cm]{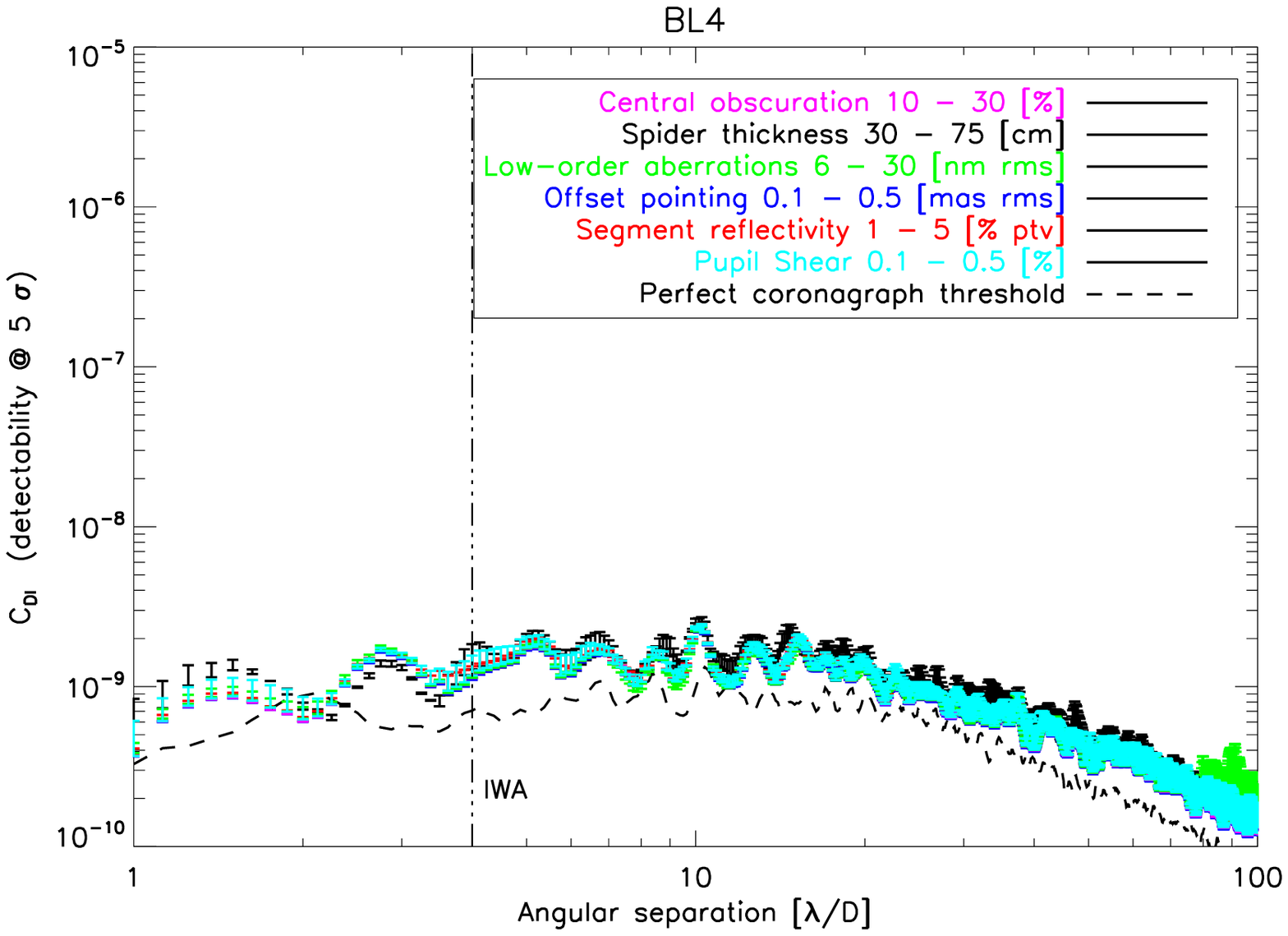}
\end{figure*} 
\begin{figure*}[!ht]
\centering
\includegraphics[width=7.65cm]{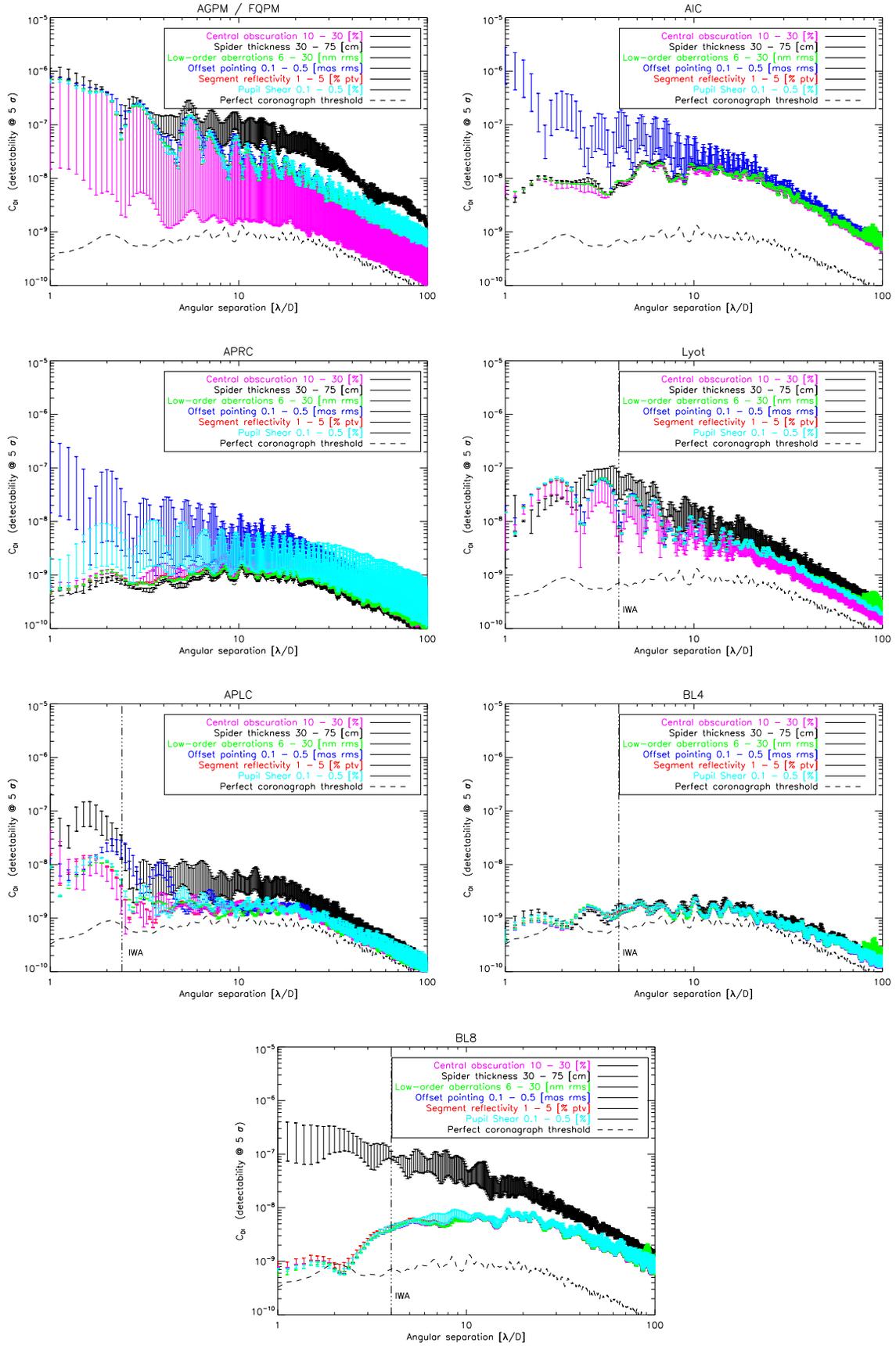}
\caption{Contrast profile at $5\sigma$, $C_{DI}(\rho)$, for the different parameters and for each coronagraph. Error bars indicates the amplitude of the contrast variation. The dashed line stands for the ideal case and was obtained for static aberrations $\delta_c$ = 10 nm rms, $\delta_{nc}$ = 0.3 nm rms. } 


\label{detecDI}
\end{figure*} 
   
\subsection{DI simulation}
\label{sec:DI}
As explained in section \ref{DIsimul} we adopted an arbitrary amount of static and differential aberrations to carry out the same analysis as in section \ref{sec:XAO2} but at a much higher level compatible with exoplanet detection. Instead of an azimuthally averaged contrast, Figure \ref{detecDI} plots the 5$\sigma$ detection level (Eq. \ref{D}). 

First of all the parameters that limit the contrast of a given coronagraph are the same except that 
at large angular distances the AO halo is no longer dominant and then the dispersion of parameters is much larger.
\smallskip

\noindent \textit{AGPM / FQPM} -- Like in previous section, a clear limit is seen at 3$\lambda/D$ between a central obscuration limited regime and a spider diffraction limited regime. Also, beyond 20$\lambda/D$, the impact of the pupil shear becomes predominant. 
The level of detectability is rather flat between 4 and 10$\lambda/D$ achieving $2.10^{-7}$.
\smallskip
 
\noindent \textit{AIC} -- Again, all symmetrical defects are quite small compared to the pointing errors. 
 At $4\lambda/D$ the performance is similar to that of the AGPM/FQPM and improves to $7.10^{-8}$ at 10$\lambda/D$ although being far from the ideal model. 
\smallskip

\noindent \textit{APRC} -- The separation between pointing errors and pupil shear limited regime has moved from 10 to 20$\lambda/D$ with respect to Fig. \ref{coroXAO}. At $4\lambda/D$ the detectability reaches $3.10^{-8}$ while at 10$\lambda/D$ is improves to $7.10^{-9}$. A gain of 1 order of magnitude is reached compared to AGPM/FQPM and AIC concepts. 
\smallskip

\noindent \textit{Lyot} -- The spider diffraction still dominates the contrast which reaches $1.10^{-7}$ at $4\lambda/D$ and improves by a factor of 10 at 10$\lambda/D$.
Considering its simplicity, the Lyot coronagraph is suitable for ELTs. 
\smallskip

\noindent \textit{APLC} -- It features one of the best detectability level with the BL4. In contrary to Fig. \ref{coroXAO}, it is dominated by the spider diffraction (and pupil shear at very large angular separation, i.e after 50$\lambda/D$) but  achieves at $4\lambda/D$ level of $1.10^{-8}$ and $8.10^{-9}$ at 10$\lambda/D$.  
\smallskip

\noindent \textit{BL4} -- Very high contrast can be achieved close to the limit imposed by static aberrations. 
The sensitivities to the parameters are rather small. 
The level of detection is identical at 4 and 10$\lambda/D$: $2.10^{-9}$.
\smallskip

\noindent \textit{BL8} -- For the same reasons expressed in Sec. \ref{sec:XAO}, BL8 is not as efficient as BL4. Up to 50$\lambda/D$ the dominating parameter is the spider diffraction, and at larger angular separations the pupil shear is dominating.
Compared to BL4, the performance degrades by about 2 orders of magnitude in the middle range of frequencies and about 1 order of magnitude at very large angular separations. At $4\lambda/D$ the detectability reaches $1.10^{-7}$ while at 10$\lambda/D$ it improves to $7.10^{-8}$. 
\smallskip

\section{Interpretations of results}
\label{sec:study2}

When considering the XAO halo, most coronagraphs (except BL8) provide roughly the same performance at mid angular radii (Fig. \ref{Perf1}). Throughput  consideration and sensitivity to parameters are then mandatory to make a difference.
In this section we summarize the most important results of the study we carried out.

For each coronagraph, the sensitivity to parameters propagates similarly between the raw coronagrahic images (XAO) and Differential images. The limiting factors are the same at these two contrast levels. 

As for band-limited, increasing the order of the mask (beyond the fourth order) is not advantageous since the Lyot stop throughput is severely restricted by the central obscuration and spiders. The Lyot stop throughput places a limit on the order of the mask that can be implemented on an ELT. Fourth order are preferable to eighth (or higher) order. This result confirms the one obtained by \citet{2007ApJ...661.1323C} where they compared Lyot type coronagraphs combined with AO system using a filled circular pupil. As already mentioned here above, the BL8 is very efficient for perfect optics but its interest is questionable in the context of ELTs. 

Coronagraphs with small IWA (AGPM/FQPM, AIC) are not able to reach the ultimate level sets by static aberrations. This is either a result of a high sensitivity to pointing errors (AIC) or an effect of the large residual amount of diffracted light by the pupil central obscuration which is not sufficiently suppressed (AGPM/FQPM). We note that solution exist to improve the peak suppression and pointing error sensitivity in the precise case of phase mask, such as the combination of a small Lyot mask placed in the center of the phase mask. A trade-off analysis would be mandatory to select the diameter of this additional Lyot mask.

For all coronagraphs, the signature of the central obscuration appears at the lowest contrast level but still can be a limitation. For instance, with the AGPM/FQPM, the other aberrations are pinned to the contrast level imposed by the central obscuration at small angular distances. Also, pointing errors and spider diffraction are critical for most coronagraph concepts (AGPM/FQPM, AIC, APRC, Lyot, BL8)

Among the concepts we have studied here, some are able to provide good and homogenous performances, namely the APRC, the APLC, the Lyot and the BL4. 

To further improve the contrast level, the main effort will have to be made on the pointing errors (telescope vibrations and stability of the XAO environment) and on the pupil shear (alignment issue). The impact of the spider diffraction shows either the importance of a coronagraph that is not sensitive to this effect (APRC, BL4, AIC), or the necessity of a specific system that can remove their impacts \citep{2006A&A...451..363A}.

Achieving a deep contrast imposes a concept of coronagraph which can accommodate some telescope characteristics while preserving a reasonable throughput. Amplitude concepts like the APLC and the BL4 appear the most suited in that case. The APLC is foreseen as the baseline design for EPICS and independent studies have shown that it is more suited to focal plane wavefront correction, a mandatory technique for EPICS. 
In the next Section, impact of the design (IWA) on these concepts and throughput considerations will be addressed. 

Finally, the simulation in Fig. \ref{detecDI} allows us to put a specification to each parameter of the simulation (within the range of values we considered) corresponding to the best contrast achievable with a given coronagraph (presented in Table. \ref{END}). This ultimate contrast level is driven in most cases by the central obscuration that we took equal to 30$\%$ in this analysis.
A coronagraph that potentially reaches high contrasts close to the level imposed by static aberrations also requires a more severe constraints on the parameters while conversely, specifications can be relaxed for a less efficient coronagraph.

\begin{center}
\begin{table*}
\centering
\caption{Preliminary parameters specification to reach best performance (set by 30$\%$ central obscuration) for each coronagraph within the simulated space parameters. Results are based on DI simulations for 30$\%$ central obscuration configuration. 
It is assumed that within specifications coronagraphs do not delivered the same detectability.}

\label{END}
\begin{tabular}{l|c|c|c|c|c}
\hline \hline 
 &\multicolumn{5}{|c}{Acceptable parameter error values} \\
\cline{2-6}
Coronagraph type & Spider [cm] &  \multicolumn{2}{c|}{Segment error}& Offset pointing [mas rms] & Pupil shear [$\%$] \\ 
\cline{3-4}
&  & Phase [nm rms] & Reflectivity [$\%$ ptv] &    &  \\
\hline
AGPM/FQPM & 30 & 30 & 5 & 0.5 & 0.5 \\
AIC & 75 & 30 & 5 & 0.2 & - \\
APRC & 75 & 30 & 5 & 0.2 & 0.3 \\
Lyot & 45 & 30 & 5 & 0.5 & 0.5 \\
APLC & 45 & 30 & 5 & 0.5 & 0.5 \\
BL4 & 60 & 30 & 5 & 0.5 & 0.5 \\
BL8 & 30 & 30 & 5 & 0.5 & 0.2 \\
\hline
\end{tabular}
\end{table*}
\end{center}

\begin{figure*}[!ht]
\includegraphics[width=9cm]{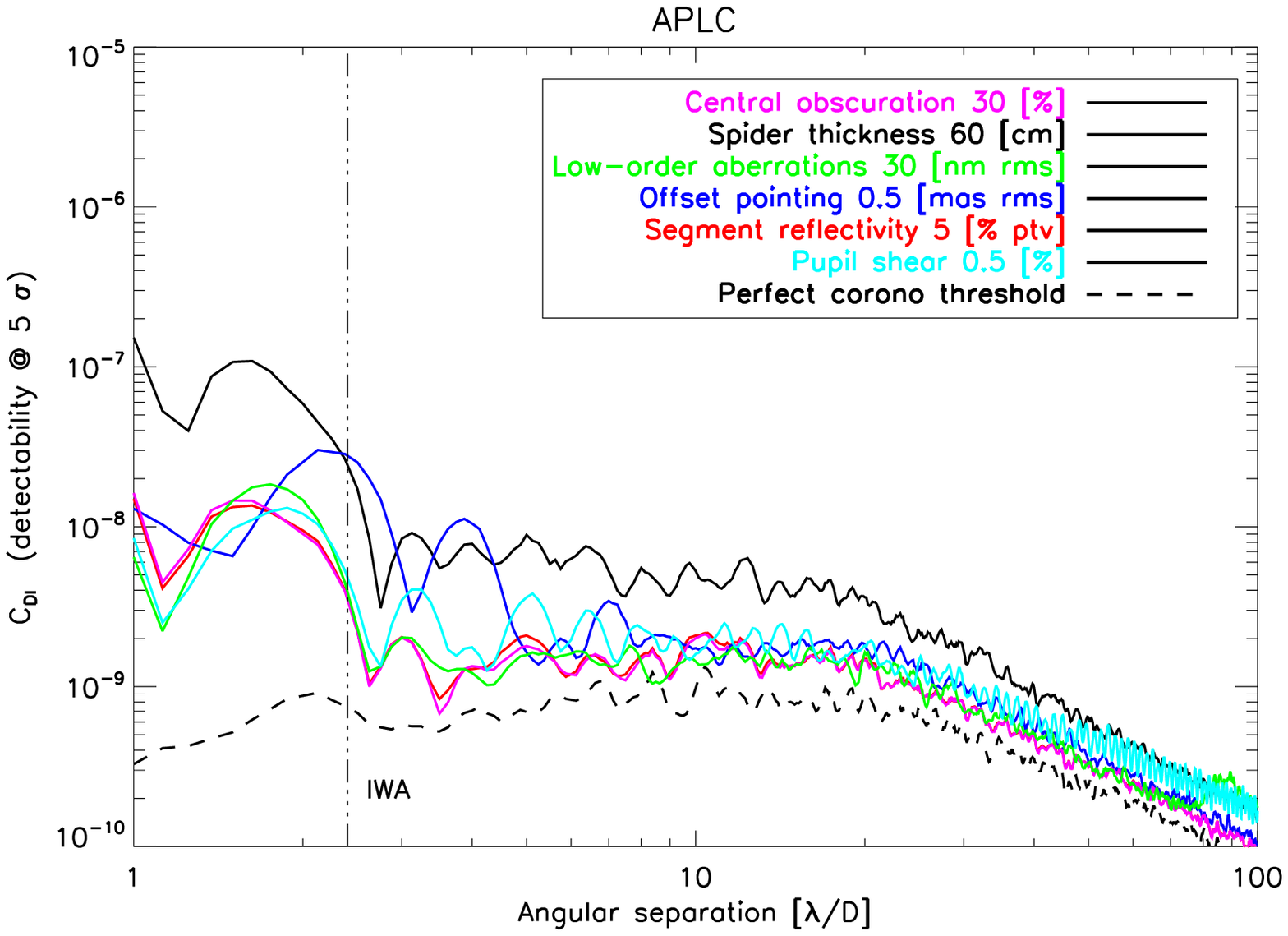}
\includegraphics[width=9cm]{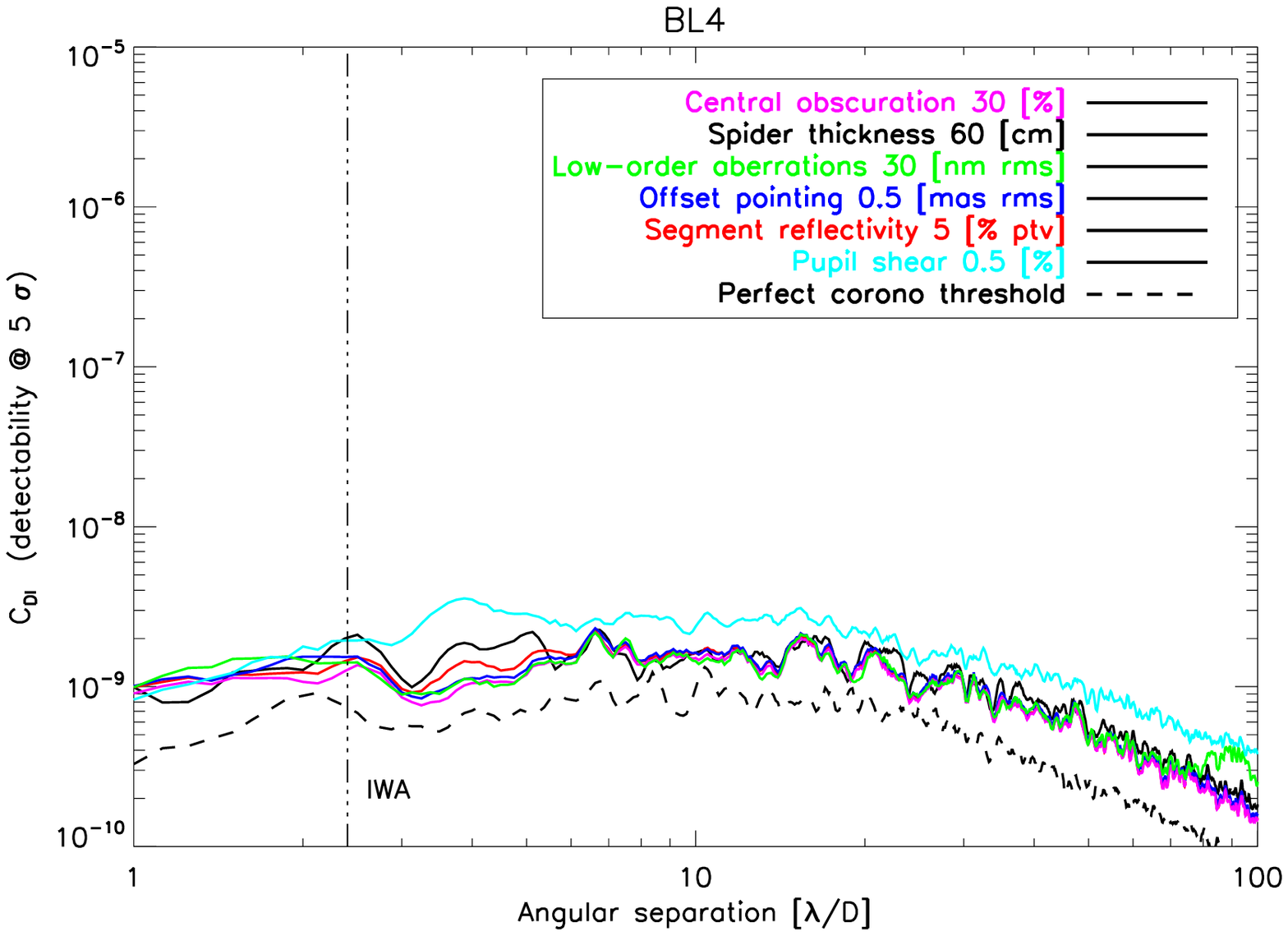}
\caption{$C_{DI}$ ($5\sigma$ detectability ($\delta_c$ = 10 nm rms, $\delta_{nc}$ = 0.3 nm rms)) vs. error sources for APLC (left) and BL4 (right) optimized for IWA = 2.4$\lambda/D$}
\label{comp}
\end{figure*} 

\section{Further comparisons of APLC and BL4}
\label{compAPLCBL4}
In previous sections, the APLC focal mask diameter (4.7$\lambda/D$, IWA=2.4$\lambda/D$) results from an optimization based on the size of the central obscuration while the mask of the BL4 is optimized for an IWA of 4$\lambda/D$. This sections compares these two designs in more details for a similar IWA.

Figure \ref{comp} presents a DI simulation when both the APLC and the BL4 are optimized for IWA=2.4$\lambda/D$ For that, the BL4 was re-optimized ($\epsilon = 0.33$, $\mathscr{T} = 12.5\%$) while APLC has remain the same ($\mathscr{T} = 54.5\%$). Here, we only present the worst case corresponding to the largest values of parameters (except for the spider thickness, sets to 60cm, E-ELT as baseline). The net effect of a smaller IWA for the BL4 is a less transmissive pupil stop, and as a result pupil shear becomes the dominant effect. From 2.4 to 20$\lambda/D$, the BL4 has a lower sensitivity to parameters, and beyond 20$\lambda/D$ both are quite comparable.

However, if we assume a comparable system transmission for these coronagraphs, the APLC will then deliver a better performance. This can be done either with a more aggressive APLC pupil stop and hence the achievable contrast is increased or conversely with a more transmissive BL4 stop. 
Even if the performance of the BL4 with a small IWA is close to that of a perfect coronagraph, its interest is questionable since the transmission is a factor of 4 lower than that of the APLC. 

Same analysis was performed for an IWA of 4$\lambda/D$. In that case the APLC has been re-optimized to 7.5$\lambda/D$ ($\mathscr{T} = 50.0\%$) while the BL4 is identical to previous sections  ($\epsilon = 0.21$, $\mathscr{T} = 22.4\%$). In such a case, the transmission is still favorable for APLC by a factor 2. Conclusions on contrast performance are identical than in the previous case.

The interest of the BL4 for ELTs would be deserved either to very bright objects or to large IWA to relax system transmission but in that case further analysis would be needed to compare its performance to that of the Lyot mask. 

\section{Conclusion}
\label{sec:conclusion}

The objective of this paper is to assess the impact of system parameters on several coronagraph concepts and to start a first order comparison in the context of ELTs. We have selected a few coronagraphs (or families) and we evaluate the behavior of the delivered contrast with respect to the main sources of degradations that occur in a coronagraphic telescope at two levels of contrast when:
\begin{itemize}
\item considering the residuals of an XAO system. 
\item a calibration of this halo is performed by the use of a differential imaging system. In that case, the residuals are set by the static aberrations.
\end{itemize}

The contrast plots that are presented in this paper are preliminary in the sense that we have considered a simple model of image formation with a limited number of parameters of which most are not yet fully defined. 
It is understood that a detailed study would involve signal to noise ratio estimation considering different type of astrophysical objects as it was done for SPHERE/VLT \citep{2005sf2a.conf..223B, 2006dies.conf..519B}. A parallel analysis has been initiated for the EPICS project  \citep{2007lyot.confE..43V}. We also note that some coronagraph concepts analyzed through this study can deliver better performance when implemented in cascade \citep{2004SPIE.5490..456A, 2007lyot.confE..25B}. 
Performance resulting from these configurations in the precise case of ELTs must be investigated further as already started for EPICS. 
Involving a wide number of existing coronagraph designs is mandatory as well (Phase Induced Amplitude Apodization Coronagraph \citep{2004AAS...20513501G}, Checkerboard-Mask Coronagraphs \citep{2004ApJ...615..555V}, for instance).

The study presented in this paper (preliminary results of system level specification are shown in Table. \ref{END}) is then one step toward this ultimate goal. Under these assumptions, we can derive three categories of coronagraphs: 
\begin{itemize}
\item those adapted for short angular separations, but conversely sensitive to pointing errors: AGPM/FQPM, AIC, APRC.  In that case, the APRC delivers the more robust performance since it is less sensitive to system parameters.
\item those adapted for intermediate angular separations: APLC and Lyots where the APLC has the advantage to provide better performance with smaller IWA and low sensitivities to system parameters. 
\item those adapted for large angular separations: BL4  and APLC.
\end{itemize}

More specifically, the APLC gathers the adequate characteristics to be a baseline design in the case of ELTs. In addition, more sophisticated implementations are possible \citep{2004SPIE.5490..456A} with the goal to provide deeper contrast and/or relax IWA constraint. Potentially, it can be upgraded (although with a particular optical system) to feature a 100\% throughput (using two mirrors apodization system based on the Phase Induced Amplitude Apodization principle \citep[PIAA,][]{2003A&A...404..379G, 2004AAS...20513501G} to generate the apodizer through beam redistribution).
 
Chromatic effects can seriously drive the choice of which coronagraph to implement. Actually, amplitude concepts are again more favorable for producing low chromatic dependencies. For instance, APLC focal plane mask size can be easily re-optimized to mitigates bandwidth effects. In the same time, many programs are striving to make other concepts achromatic, as the AGPM or the multi-FQPM \citep{2005SPIE.5905..486M, 2005ApJ...633.1191M, 2007lyot.confE..25B}, achromatic and improved versions of the FQPM. 

However, to fully take advantage of a coronagraph the most demanding parameters is clearly the level of the XAO residuals and then a lot of efforts has to be made to provide very high Strehl ratios on ELTs. 

Finally, the manufacturing feasibility of coronagraphs is also a critical issue in the development of an high contrast instrument for ELTs. In that perspective, we have started to prototype several designs (APLC, FQPM, Lyot and BL, \citet{HOTcorono}) to be tested on the High Order Test-bench developed at the European Southern Observatory \citep{2006SPIE.6272E..81V, HOTbench}. Results of these technical aspects will be presented in forthcoming papers. 

\begin{acknowledgements}
P.M would like to thank Lyu Abe for helpful discussions on two mirrors apodization for APLC. This activity is supported by the European Community under its Framework Programme 6, ELT Design Study, Contract No. 011863. 
\end{acknowledgements}

\nocite{*}
\bibliography{MyBiblio}
\newpage
\clearpage
\end{document}